\documentclass[epj,nopacs]{svjour}
\pdfoutput=1
\usepackage[utf8]{inputenc}
\usepackage{caption,amsmath,amssymb,amsfonts,graphicx,cite,multirow,tikz,enumitem}
\DeclareCaptionType{Diagram}
\usepackage{pstricks,pdflscape,morefloats,algorithmicx,algpseudocode,slashed}
\usepackage[section]{algorithm}
\usepackage{booktabs}

\usepackage{color}
\usepackage{hyperref}

\makeatother

\newcommand{\hw}{Herwig~7}

\def\qb{\bar q}

\usepackage{mathtools}

\preprint{CoEPP-MN-19-04, MCNET-19-21, KA-TP-08-2019, HERWIG-2019-01, LU-TP 19-41}

\title{Spacetime Colour Reconnection in Herwig~7}

\author{Johannes Bellm\inst{1}\and
Cody B Duncan\inst{2,3}\and
Stefan Gieseke\inst{3}\and
Miroslav Myska\inst{4}\and
Andrzej Si\'odmok\inst{4,5}
}

\institute{
Theoretical Particle Physics, Department of Astronomy and Theoretical Physics, Lund University, Lund, Sweden \and
School of Physics and Astronomy, Monash University, Clayton, VIC 3800, Australia \and 
Institute for Theoretical Physics, Karlsruhe Institute of Technology, 76128 Karlsruhe, Germany \and
Czech Technical University in Prague, Brehova 7, 115 19 Prague, Czech Republic \and
The Henryk Niewodniczanski Institute of Nuclear Physics in Cracow
}

\date{\today}

\abstract{We present a model for generating spacetime coordinates
in the Monte Carlo event generator \hw, and perform colour reconnection
by minimizing a boost-invariant distance measure of the system. We compare the model
to a series of soft physics observables. We find reasonable agreement with
the data, suggesting that $pp$-collider colour reconnection
may be able to be applied in larger systems.
}

\begin{document}

\maketitle

\section{Introduction}
As the LHC reaches unprecedented levels of precision and data collection,
the playground for studying QCD effects has increased manifold. In particular,
Monte Carlo event generators 
\cite{Bahr:2008pv, Bellm:2017bvx ,Herwig7, Sjostrand:2014zea, Gleisberg:2008ta} 
provide an ideal arena for testing novel ideas
in the low-energy regime, i.e. the mechanisms of hadronization,
where non-perturbative effects have to be
phenomenologically modelled, and the underlying event.
One aspect of proton-proton collisions that is poorly understood
is exactly how multiple parton-parton interactions from the initial
scattering process interfere and interact with one another during
the hadronization stage.

Multiple parton interactions were first introduced in \cite{Sjostrand:1987su}, and
implemented in Pythia \cite{Sjostrand:2014zea}, where its importance
in hadronic collisions was highlighted beyond a doubt.
A similar physical notion was introduced in \cite{Butterworth:1996zw} and later 
implemented in Herwig++ \cite{Bahr:2008dy,Bahr:2008pv,Bahr:2009ek}, 
with some recent improvements to soft and diffractive scatterings in
\cite{Gieseke:2016fpz,Bellm:2017bvx} to \hw{}.

One such model of this interference between subcollisions in an event is
colour reconnection
\cite{Sjostrand:1993hi, Gieseke:2012ft, Bierlich:2015rha,Gieseke:2017clv,Christiansen:2015yqa},
whereby a Monte Carlo
event generator reduces some kinematic, momentum-based measure 
of the event. The physical intuition for such a mechanism is twofold:
to correct for errors in the leading-colour approximation of the 
parton shower, and
to allow multiple parton interactions, which may have been colour-connected, 
to have cross-talk.
A summary of the history of colour reconnection and the effects
of such a mechanism on precise measurements is given in
\cite{Sjostrand:2013cya}.
Colour reconnection in \hw{} first focused on reconnecting excited
$q\bar{q}$ pairs called clusters, minimizing the sum of the invariant
masses. Later work \cite{Gieseke:2017clv} expanded upon this model 
to introduce
the possibility of forming so-called baryonic 
clusters $qqq$ and $\bar{q}\bar{q}\bar{q}$
from three ordinary/mesonic clusters. Other methods have investigated
colour reconnection at the perturbative stages of event simulation or
taken inspiration from perturbative techniques
\cite{Lonnblad:1995yk,Gieseke:2018gff,Bellm:2018wwz}.

Most $pp$ event generators are developed in the energy-momentum 
framework for the various
stages of event simulation, meaning that none of the physics modelled
involves any notion of spacetime separation.
While the energy-momentum 
framework has been very successful,
there are still several issues at hand. In particular, it does not
have an adequate answer to what parts of the event are allowed
to undergo colour reconnection within a given slice of phase space, if one
thinks that colour reconnection needs to be a causal effect.
Collisions of heavy ions have shown that spacetime structure is important
in modelling where interactions start, since a jet starting at the 
edge of the quark-gluon plasma will lose far less energy to one travelling
through the centre of dense medium, a phenomenon known as jet
quenching \cite{Bjorken:1982tu,Gyulassy:1990ye,Qin:2015srf}. 
As a result, $pp$-oriented event generators
have also started to include more spacetime information, using these
coordinates for various aspects of the simulation, such as
collective hadronization effects 
\cite{Bierlich:2014xba,Bierlich:2016vgw}, and
a spacetime evolution of the parton shower \cite{Ellis:1995fi}.
Pythia recently introduced a framework for generating
spacetime coordinates \cite{Ferreres-Sole:2018vgo} 
for quantitative studies of 
Lund string fragmentation \cite{Andersson:1983ia}. The effects of
introducing spacetime coordinates have been recently studied in
dipole evolution in
$\gamma^{*}A$ collisions \cite{Bierlich:2019wld}.

As high energy and heavy ion phenomenology begin to have more
interaction with each other, an immediate question one should ask is
if the models developed in each field can be applied to the other
successfully. Without spacetime information, high energy event generators
cannot hope to be able to describe hadronization of large systems well. This
work aims to be the first steps of introducing spacetime coordinates
and using them to aid the baryonic colour reconnection model
\cite{Gieseke:2017clv}. We intend this to be a proof of concept that
will allow us to apply this hadronization model to heavy ions in later work.

The format of the article is as follows: we start by recalling
elements of modelling high energy collisions, such as the underlying
event, cluster 
hadronization, and colour reconnection models in \hw{}, in
Sec. \ref{sec:hadronizationmodel}. In Sec. \ref{sec:coords}, we describe
our method of systematically assigning coordinates to the multiple
parton interactions and the partons at the end of the shower. 
We then present our model of using this spacetime 
information to perform colour reconnection
in Sec. \ref{sec:spacetimeCR}. We briefly describe additional
modifications that have been applied in the making of 
this and related works in Sec. \ref{sec:modifications}. 
We tune our new model 
in Sec. \ref{sec:tuning} and present the results of the procedure in
Sec. \ref{sec:results}. Lastly, with Sec. \ref{sec:conclusion}, 
we summarize our model and future work.

\section{Event Simulation in \hw}
We briefly summarize the pertinent points of modelling the underlying event 
and hadronization in \hw.
\label{sec:hadronizationmodel}

\subsection{Multiple parton interactions (MPI)}
\label{sec:mpi}
Since the proton is a composite particle, when two protons collide,
there may be several parton-parton interactions, which fall into
two classes in \hw: hard and soft. Partons from hard scatters
undergo parton showering, while soft scatters do not.

For a given event, \hw{} generates a number of each type of these scatters.
The average number of interactions for a given 
impact parameter $b$ and centre of mass energy $s$ is schematically 
given by:
\begin{equation}
\langle n_{\mathrm{int}}\rangle = A(b;\mu) \sigma^{\mathrm{inc}}(s;p_{\perp}^{\mathrm{min}}) ,
\label{eq:avgn}
\end{equation}
where $\sigma^{\mathrm{inc}}$ is the inclusive cross section to produce a pair
of partons above a defined minimum transverse momentum, 
$A(b;\mu)$ is the overlap function between the two protons,
and $\mu^2$ is commonly referred to as the inverse hadron length. 
In \hw,
both the hard and soft MPI scatters have the same form for Eq. 
\ref{eq:avgn}, and indeed it is assumed that they both have the same
functional form for the overlap function, 
but with different values for $\mu^2$. Similarly, the inclusive cross 
sections are different values for hard and soft scatters.

\hw{} assumes the MPI to be independent of one another (including 
energy-momentum conservation), leading to a Poissonian probability 
distribution. Using the notation of \cite{Herwig7}, we can 
write the joint probability distribution to produce $h$ hard and 
$k$ soft scatters at a given b\footnote{We have suppressed the 
functional dependence on centre of mass energy $s$.} as:
\begin{equation}
\mathcal{P}_{h,k}(b) = \frac{\left(2\chi_h\right)^{h}}{h!}
\frac{\left(2\chi_k\right)^{k}}{k!}e^{-2(\chi_h+\chi_k)} ,
\label{eq:poissonmpi}
\end{equation}
where $2\chi_{h,k} = A(b;\mu_{h,k})\sigma^{\mathrm{inc}}_{h,k}$ is the so-called eikonal factor.
This formalism was developed in \cite{Aurenche:1991pk} and Herwig's implementation
is built on the JIMMY framework \cite{Butterworth:1996zw}.

Eq. \ref{eq:poissonmpi} is then integrated over $b$ space to produce
an exact probability to produce the corresponding number of hard and 
soft scatters in an event:
\begin{equation}
	P_{h,k} = \frac{\int\text{d}^2 b \mathcal{P}_{h,k}(b,s)}{\int\text{d}^2 b\sum_{h\geq1,k=0}^{\infty}\mathcal{P}_{h,k}(b,s)} .
	\label{eq:mpiSampling}
\end{equation}
\hw{} samples the distribution in Eq. \ref{eq:mpiSampling} probabilistically,
to obtain a number $h$ of hard scatters, and $k$ of soft scatters. 
The primary hard subprocess in Minimum Bias event generation in \hw{} is an interaction
between two valence (antiquarks) \cite{Gieseke:2012ft}, while subsequent MPI collisions
are initiated by regular $2\to 2$ QCD processes. The incoming legs are
evolved backwards to pairs of gluons extracted from the beam remnant, with the colour
topology defined in the $N_C \to \infty$ limit. The colour topology is motivated
by the leading-colour approximation used in the shower, though as discussed in \cite{Gieseke:2012ft},
this is a phenomenological choice rather than an approximation.

As \hw{}
produces each scatter, it checks the available energy and momentum in the protons.
If the protons cannot produce another
scatter, the MPI production algorithm terminates.
As a result, \hw{} typically generates a subset of the total number of scatters
sampled from Eq. \ref{eq:mpiSampling}. More details of the technicalities
involved in the implementation of MPI algorithm can be found in \cite{Bahr:2008pv}.

\subsection{Cluster model}
\label{sec:cluster}
Partons from a scattering process
are showered down to the parton shower cutoff scale, and 
the resulting colour topology
has triplets connected to anti-triplets via gluon connections.
At the hadronization scale and below, \hw{} uses the cluster 
hadronization model \cite{WEBBER1984492}, based on the pre-confinement
property of angular-ordered showers \cite{AMATI197987}.

The first step in the cluster model is to non-perturba\-tively split the gluons
into quark-antiquark pairs.
To split the gluons, \hw{} uses a kinematic map at the end of the shower to
put the gluons on-constituent-mass-shell and performs an isotropic decay.
The constituent-mass of the gluon is a non-perturbative parameter of \hw{} hadronization
model.

Nearest quark-antiquark neighbours in colour space,
which are typically nearest neighbours in momentum space
due to pre-confinement, are then
collected into colourless, excited quark-antiquark pairs, i.e. clusters.
From there, the clusters undergo colour reconnection.

\subsection{Colour reconnection}
\label{sec:cr}
Clusters typically connect partons from the same multiple 
parton interaction scattering. 
Colour reconnection alters the colour topology
of the event, and allows the different MPI to interact with one another
at the hadronization level.

As mentioned in Sec. \ref{sec:mpi}, \hw{} chooses the leading-colour
topology for the additional scatters, thus they are colour-connected to the beam
remnant and other subprocesses. As noted in \cite{Gieseke:2012ft}, colour reconnection
is a required part of hadronization modelling in hadron collisions since
the leading-colour approximation performs significantly worse in
non-perturbative parts of the event generation.

Colour reconnection aims to minimize a given measure of the event,
typically momentum-based. \hw{} has a variety of colour reconnection
algorithms \cite{Gieseke:2012ft,Gieseke:2017clv}, namely:
\begin{itemize}
    \item Plain,
    \item Statistical/Metropolis,
    \item Baryonic.
\end{itemize}
The plain colour reconnection model locally minimizes pair-wise
cluster invariant masses:
\begin{equation}
    m_{q\bar{q}}^2 = \left(p_q + p_{\bar{q}}\right)^2 .
\end{equation}
The criteria for two clusters to
undergo colour reconnection and swap partners is:
\begin{equation}
    m_{q\bar{q}'} + m_{q'\bar{q}} < m_{q\bar{q}} + m_{q'\bar{q}'} .
    \label{eq:massMin}
\end{equation}
If a pairing reduces the invariant mass, it is allowed to reconnect with a
flat probabilistic weight, typically tuned to LHC data, 
while ensuring that the model doesn't adversely affect LEP
simulations.
Baryonic colour reconnection was recently implemented in \hw{} 
\cite{Gieseke:2017clv}, and it uses a more sophisticated
algorithm. For each cluster in the event, the algorithm
searches for other clusters which 
occupy the same neighbourhood in rapidity-space. It searches for two types
of candidate clusters for reconnection: baryonic, and 
(ordinary) mesonic.

In the baryonic case, given a cluster A, transform the momenta of all other
clusters to the rest frame of A, and search for two other clusters that
have the same orientation of quark axis in rapidity space.
It then chooses the pair of candidate clusters which 
have the largest rapidity span in this frame.
If the reconnection is accepted, 
the quarks are then collected into a three-component cluster, called
a baryonic cluster, and similarly the antiquarks are collected into
an anti-baryonic cluster.

In the mesonic case, if the candidate cluster B with the largest 
rapidity span
has a quark axis oriented in the opposite direction to cluster A, reconnect
$q_A\bar{q}_B$ and $q_B\bar{q}_A$, in much the same manner as the plain
colour reconnection model.
For both types of cases in baryonic colour reconnection, 
the probabilities for reconnection are given
by two different flat weights, $p_{\text{M,reco}}$ and $p_{\text{B,reco}}$.

While the statistical colour reconnection model is outside the scope
of this paper, we mention that it aims to minimize mass, much like
the plain model, but it allows reconnection to increase the mass of the
system with a suppressed probability, and is based on the simulated annealing
optimization algorithm \cite{Kirkpatrick671}.  

In all cases, colour reconnection qualitatively 
aligns colours between partons that move into the same direction 
such that the 
multiplicity of particles produced in between them is reduced and 
the produced particles carry more momentum on average. 

\section{Spacetime Coordinate Generation}
\label{sec:coords}
We present the two parts of how our model systematically generates coordinates
for the multiple parton interaction scattering centres and the hadronization
stage. We argue that these are the two stages of event generation
that are most impactful on spacetime coordinates.

\subsection{MPI coordinate generation algorithm}
To obtain an intelligent and relevant 
value for the impact parameter, the MPI coordinate generator takes 
the produced values for $h, k$ in Eq. \ref{eq:mpiSampling}
and stochastically samples the distribution of Eq. 
\ref{eq:poissonmpi}, vis-a-vis a veto algorithm. Thus, the produced 
$b$, when the number of events tends to infinity, will be the correct 
distribution for a given set of $h$ and $k$. 

\begin{figure*}
\centering
\includegraphics[width=0.8\textwidth]{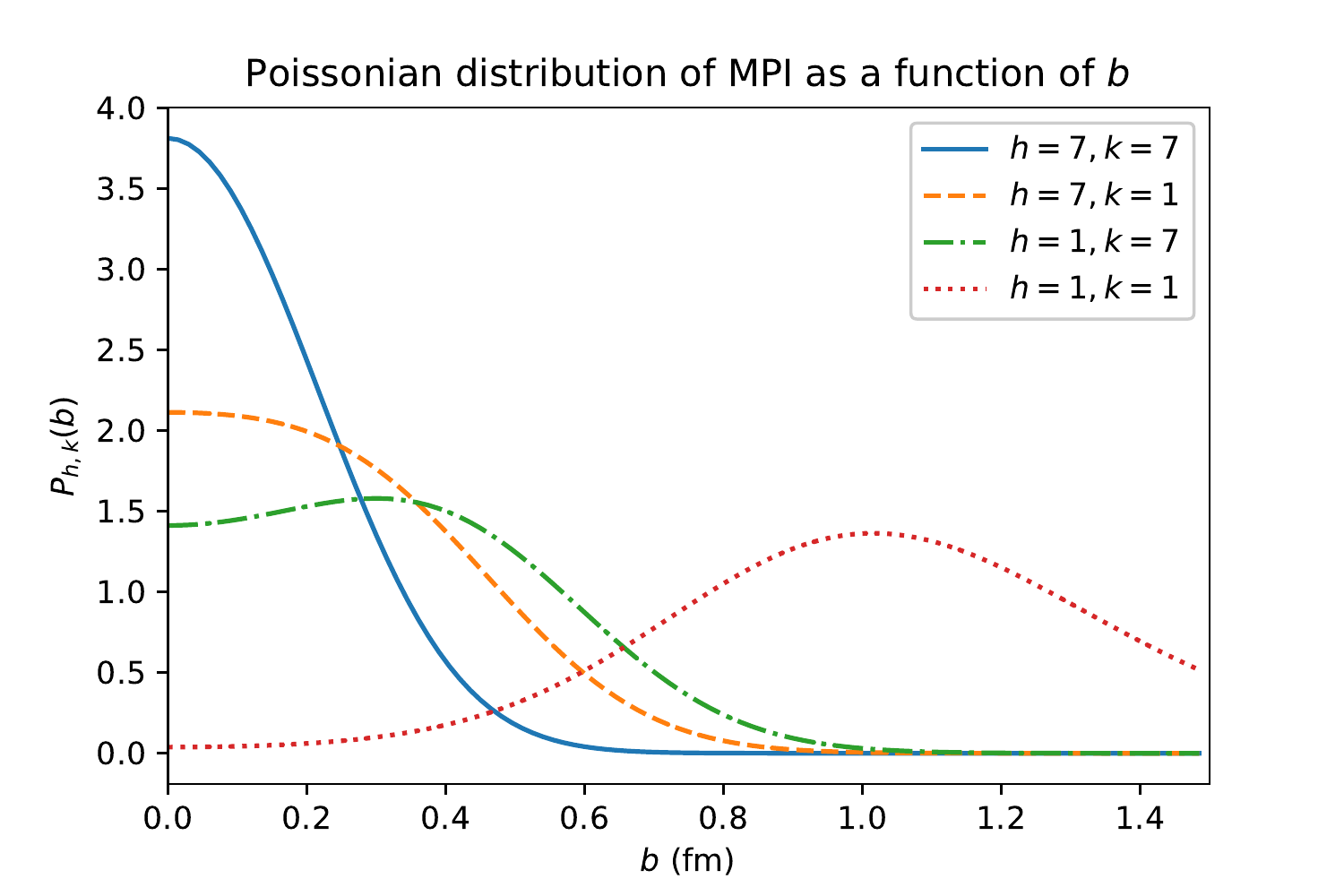}

\caption{ Joint Poissonian distribution $\mathcal{P}_{h,k}(b)$, 
as a function
of impact parameter $b$, for a number of $h$ hard scatters
and $k$ soft scatters. We have picked one large (7) and one 
small (1) value, and show the various combinations. The more 
collisions that occur, the more likely the collision is to be
central. Keeping the number of interactions fixed but having more soft
interactions makes the distribution have a broader tail.
We have used the following fixed values for the normalized distributions: 
$\sigma^{\mathrm{inc}}_{\mathrm{hard}} = 83~\mathrm{mb}$,
$\sigma^{\mathrm{inc}}_{\mathrm{soft}} = 127~\mathrm{mb}$, 
$\mu^2_{\mathrm{hard}} = 0.71~\mathrm{GeV}^2$, and 
$\mu^2_{\mathrm{soft}} = 0.52~\mathrm{GeV}^2$. These distributions are normalized
independently to unit area.}
\label{fig:jointpoissonian}
\end{figure*}

As shown in 
Fig. \ref{fig:jointpoissonian}, the joint Poissonian behaves as we expect.
The more scatters that \hw{} produces, the more likely it is that
the sampled $b$ will be central, while having more soft scatters for a fixed
number of total scatters makes the distribution have a broader tail.
In this work we will be using the Bessel proton profile, meaning
that the overlap function is a Bessel function of the third kind:
\begin{equation}
  \label{eq:overlap}
  A(b;\mu) = \frac{\mu^2}{96\pi}(\mu b)^3K_3(\mu b) .
\end{equation}

It should be noted that the results of the sampling should not be 
surprising. At large numbers of interactions, the sampled impact 
parameters tend to be closer to 0, since a larger than 
average number of interactions requires a more central collision.
Once $b$ is determined for a given event,
we set the incoming beam positions to be at
$(\pm b/2,0,0,0)$, i.e. aligned along the $x$-axis, for simplicity.

The overlap function $A(b;\mu)$ in Eq. \ref{eq:overlap} is generated
by the convolution of the two protons' form factors, $G(b;\mu)$:
\begin{equation}
    \label{eq:convolution}
    A(b) = \int \mathrm{d}^2 b' G(b') G(b-b') ,
\end{equation}
where we have suppressed the dependence on $\mu$ for clarity.
The overlap function governs the density of MPI scattering centres
in the transverse plane for a given offset between the protons.

To obtain the MPI centre positions, we sample
the integrand of Eq. \ref{eq:convolution}.
We generate $h$ hard scatters, and $k$ soft scatters, using two
different $\mu^2$ values for the hard and soft interactions.
As a result, hard scatters are slightly more concentrated in the centre
of the transverse plane, while soft scatters have a longer tail.

\begin{figure*}[th]
\centering
\includegraphics[width=0.49\textwidth]{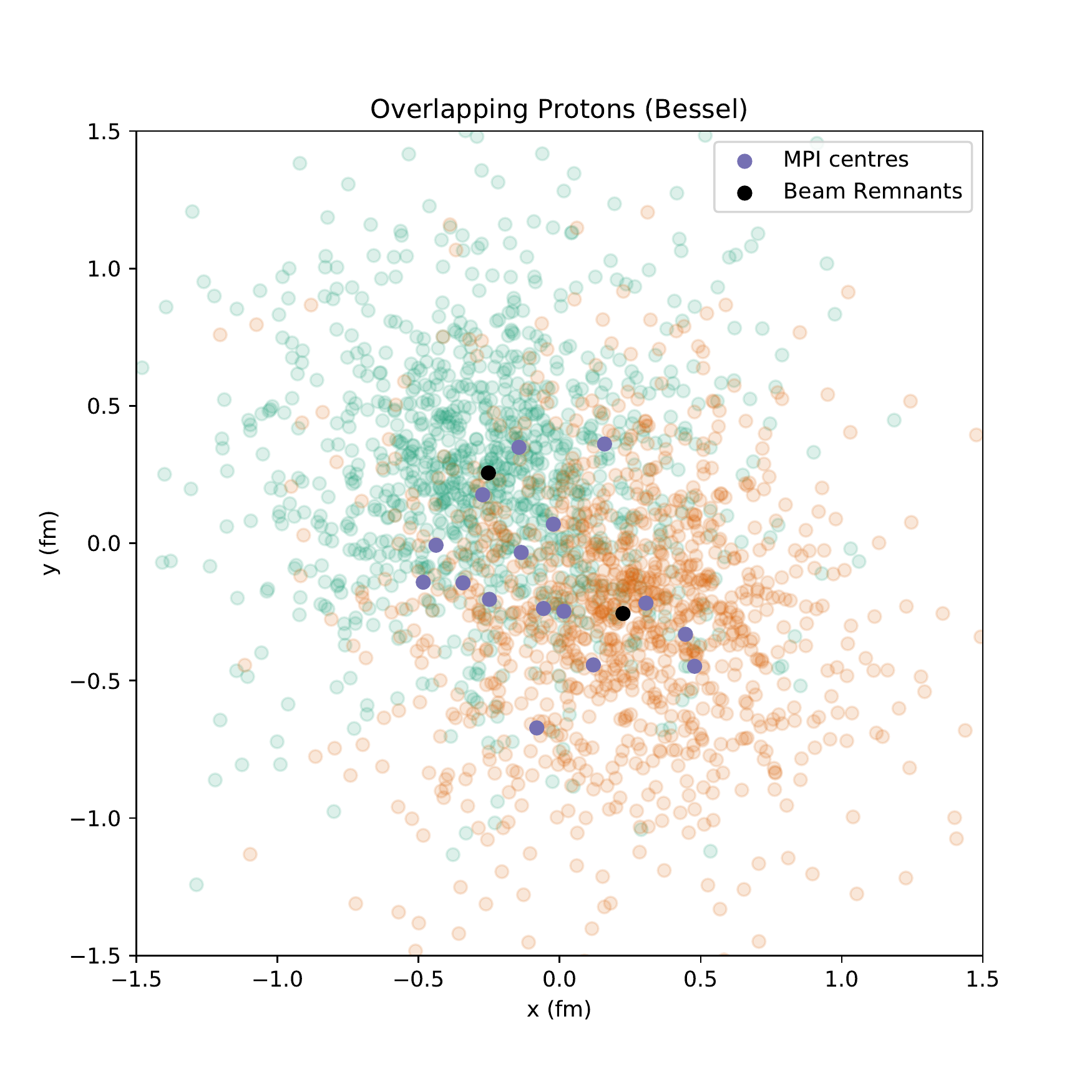}
\includegraphics[width=0.49\textwidth]{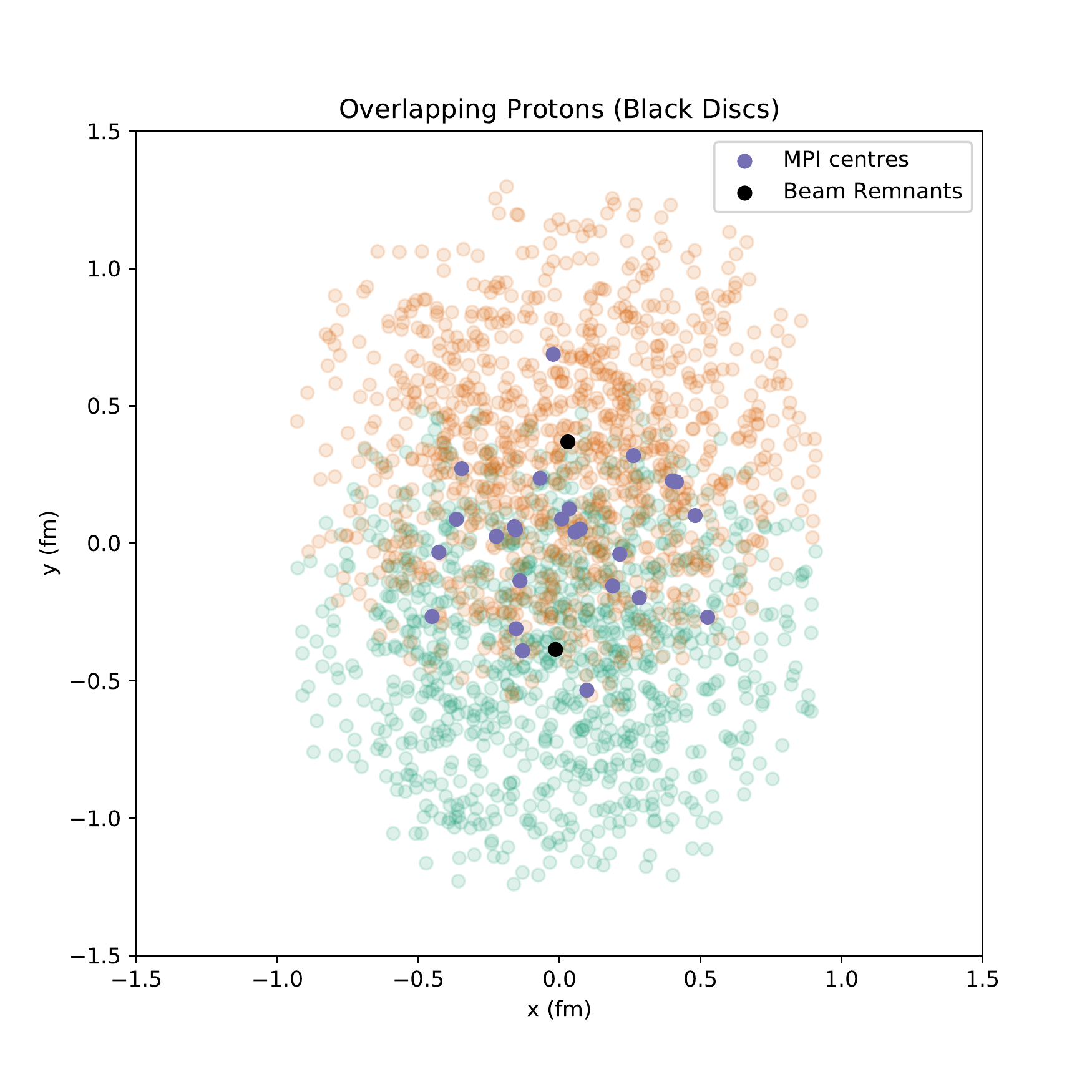}
\caption{ Result of MPI Coordinate Generator algorithm with 
the Bessel proton profile (left), and an example for a uniform (black disc)
proton profile (right). Green and orange points are partons 
sampled in a given proton, mauve points are accepted 
MPI collision centres, and black are the beam remnants.
}
\label{fig:mpicoords}
\end{figure*}

Once these points have been generated, all coordinates
including the proton positions get the same random global rotation in the
transverse plane.
The beam remnants receive the sampled proton positions.
A schematic diagram 
of the results of the MPI coordinate generation algorithm is shown 
in Fig. \ref{fig:mpicoords}. The overlap need not necessarily be a Bessel
function, and we have included the results of the MPI coordinate generation
for a uniform proton profile in Fig. \ref{fig:mpicoords}. For this type of
proton profile, MPI centres can only be situated in the overlap.
However, for the rest of the
paper, we will work with the Bessel function profile.

\subsection{Tracing spacetime during parton showers}
\label{sec:propagation}
The spacetime structure of the parton-shower evolution was already 
considered in the early paper on QCD cascades by Fox and Wolfram (see Fig. 1 of \cite{Fox:1979ag}).
Later the spacetime evolution of the parton shower was introduced, for example, to study jets in hadronic $e^+e^-$ 
events at LEP~\cite{Ellis:1995dq} and in deep-inelastic ep scattering~\cite{Ellis:1996nv}.
Very recently in a publication on the space--time structure of hadronization 
in the Lund Model~\cite{Ferreres-Sole:2018vgo} the authors mention that a sensible spacetime 
picture of parton-shower evolution would introduce some spacetime offsets
to their model. However, the authors assumed that the offsets 
are most likely small in their case and therefore neglected them in their studies. 

In the following section, we will investigate in 
more detail how the parton shower affects the spacetime structure 
of an event as implemented in the family of \hw{} generators.
Referring to~\cite{Corcella:2000bw} (Section 3.8) for details, we briefly recall the 
essential concepts of the \hw{} spacetime model.
It should be noted that there are two major parton shower options in
Herwig, namely the angular-ordered shower \cite{Gieseke:2003rz} 
and the dipole shower \cite{Platzer:2009jq}.
For this work, we will focus on the angular-ordered shower, and its use
of virtuality as an evolution variable.

The mean lifetime $\tau$ of a parton in its own rest frame, during
the parton shower evolution, is calculated in a similar 
manner as for particles decays, i.e. taking into account its natural
width $\Gamma$ and virtuality $q^2$:
 \begin{equation}
     \tau(q^2) = \frac{\hbar\sqrt{q^2}}{\sqrt{\left(q^2 - M^2\right)^2 +\left(\frac{\Gamma q^2}{M}\right)^2}} ,
     \label{eq:spacetimePropagation}
 \end{equation}
Eq. \ref{eq:spacetimePropagation} interpolates between 
the lifetime for an on-mass shell parton
$\tau(q^2 = M^2) = \hbar/\Gamma$, and for a highly virtual (i.e.
off-mass shell) parton $\tau(q^2 \gg M^2) = \hbar/\sqrt{q^2}$.
We note that the mean lifetime in Eq. \ref{eq:spacetimePropagation}
is equivalent to the standard notion
of formation time used in heavy ion phenomenology as well as in general
jet quenching research
\cite{Baier:2000mf,Casalderrey-Solana2016,Blaizot:2019muz,Dominguez:2019ges}. 
We show the equivalence in App. \ref{app:time}.
\footnote{The authors are grateful for Gavin Salam's notes on the notion
of formation time for massless soft and collinear gluons.}

Once a lifetime is calculated according to Eq. \ref{eq:spacetimePropagation},
the parton decays according to an exponential
decay law, with a rest-frame decay time $t^*$:
\begin{equation}
    P_{\text{decay}}(t < t^*) = 1 - \exp\left(-\frac{t^*}{\tau}\right) .
    \label{eq:decaylaw}
\end{equation}
After sampling a rest-frame decay time, this time can be 
converted to the lab-frame decay time $t$, and a distance 
travelled in the lab-frame, $\vec{d}$:
\begin{equation}
    t = \gamma t^* , \vec{d} = \vec{\beta} \gamma t^* ,
    \label{eq:labCoords}
\end{equation}
where $\gamma$ and $\vec{\beta}$ are the usual Lorentz factors.

Very light quarks and gluons with a small natural width
may travel unphysically large distances according to Eq. \ref{eq:spacetimePropagation}
in the final steps of the parton shower.
Similarly, there are issues with assigning particles with no 
well-defined width spacetime
coordinates in the above manner.
In order to counter this issue, a minimum width $\Gamma=\nu^2/M$
is introduced, where $\nu^2$ (GeV$^2$) is a free parameter of the 
order of lower limit of parton's virtuality.
This is essentially the spacetime equivalent of a shower 
$Q^2 \approx \Lambda_{\mathrm{QCD}}^2$ cutoff scale.
The daughters of the parton splitting are then given the
starting coordinates defined by Eq. \ref{eq:labCoords}.
We note that the above considerations are, in our model, a
phenomenological model of the spacetime structure of an event,
which arise during the initial collision of the protons, and the
subsequent perturbative evolution of the event.

In order to study the size of the parton-shower spacetime effects, we will first
consider the distance that each parton propagates during the shower. The
distance that we are interested in is the difference between a given parton's
production and decay vertex, $L$:
\begin{equation}
    L = \sqrt{\left(d_{\rm decay} - d_{\rm prod}\right)^2} ,
\end{equation}
where $d \equiv d^{\mu} = (t,x,y,z)$ is the position of a parton relative to the
centre of the collision, i.e. the origin.
However, since the MPI smearing discussed in the previous section affects 
only the transverse plane we will also consider
transverse distance, constructed from the transverse components
of the above vertices, $r = \sqrt{\Delta x^2+\Delta y^2}$.

\begin{figure*}[th]
\centering
\includegraphics[width=0.49\textwidth]{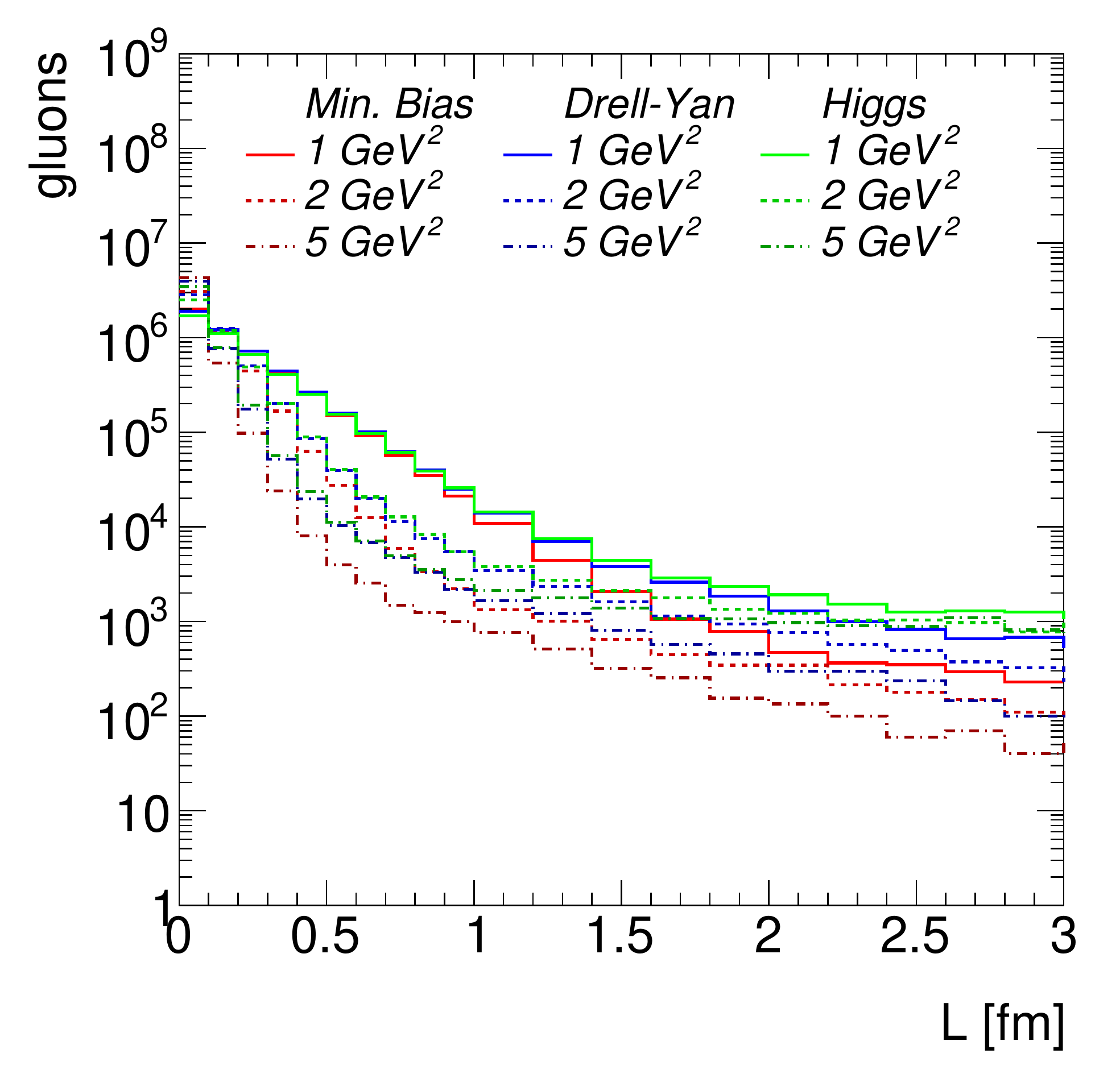}
\includegraphics[width=0.49\textwidth]{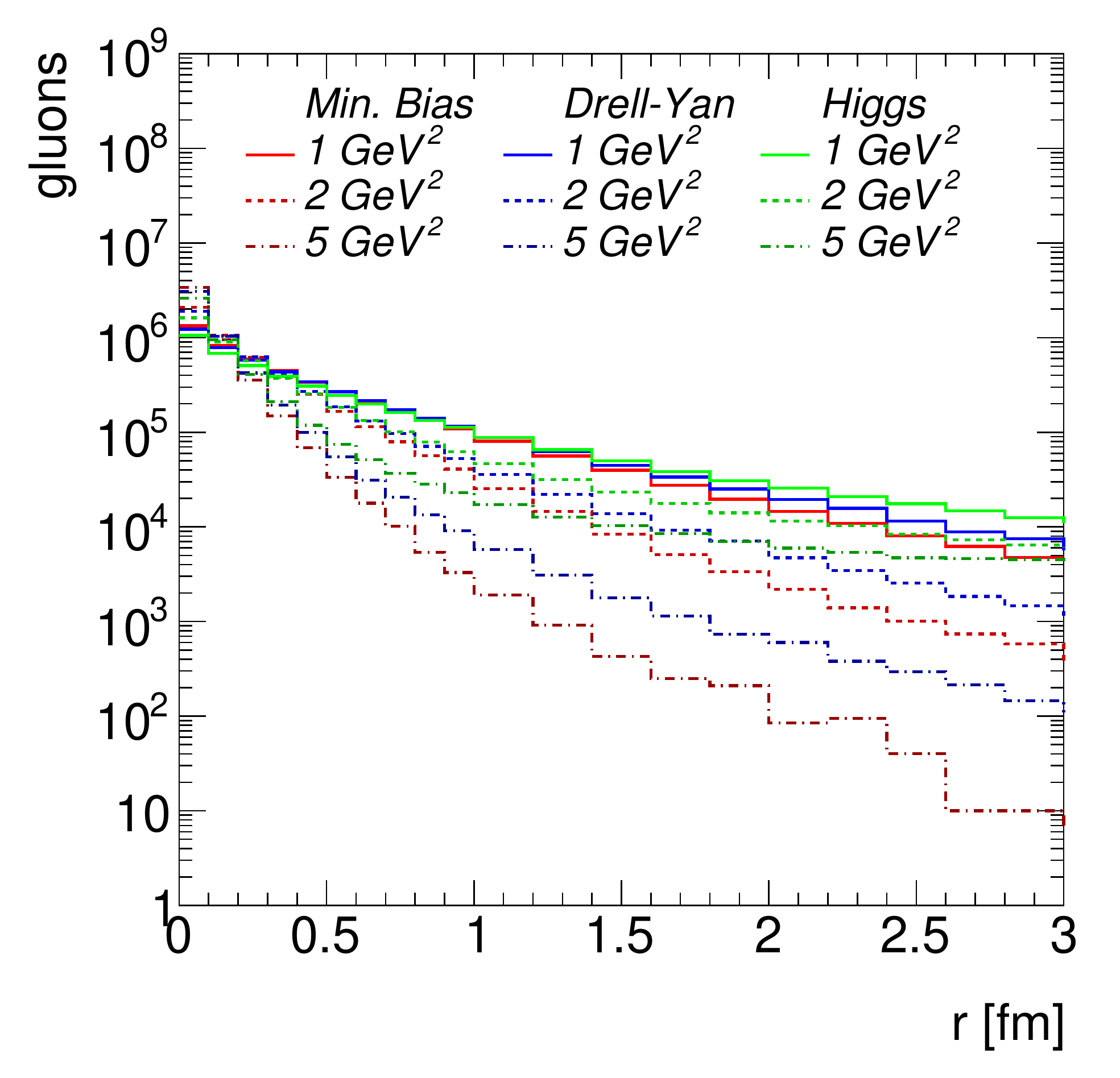}
\caption{
The total Lorentz-invariant
distance $L$ (left panel) and transverse distance $r$ (right panel) traveled by the
gluons at the last step of the parton shower evolution for three different processes: Minimum Bias,
Drell-Yan and Higgs-boson production at the LHC at the centre-of-mass energy $7$~TeV.
The simulation was performed using default version of \hw{} using 
three different values of $\nu^2$: $1$, $2$ and $5$~GeV$^2$.}
\label{fig:Lr}
\end{figure*}

In Fig.~\ref{fig:Lr} we show the Lorentz-invariant
distance $L$ (left panel) and transverse distance (right panel) traveled by the
gluons at the last step of the parton shower evolution for three different processes: Minimum Bias,
Drell-Yan and Higgs-boson production at the LHC at the collision energy $7$~TeV.
The simulation was performed using default version of \hw{} with 
three different values of $\nu^2$: $1$, $2$ and $5$~GeV$^2$.
We see that most of the partons reach fermi-scale distances which are comparable to
the size of the MPI coordinate generation, as shown in Fig.~\ref{fig:mpicoords}. 
Therefore, it is important to
take the parton shower effects into account. We also see that in soft Minimum Bias 
processes the partons travel shorter distances, as expected 
since there is less parton-shower activity in these types of events
than in the two other processes.
Finally we see that the results, and especially the long distance tails of the distributions, 
are strongly dependent on the scale $\nu^2$. This indicates that the
furthest distances are traveled by partons in the final step of the evolution.

\begin{figure*}[th]
\centering
\includegraphics[width=0.49\textwidth]{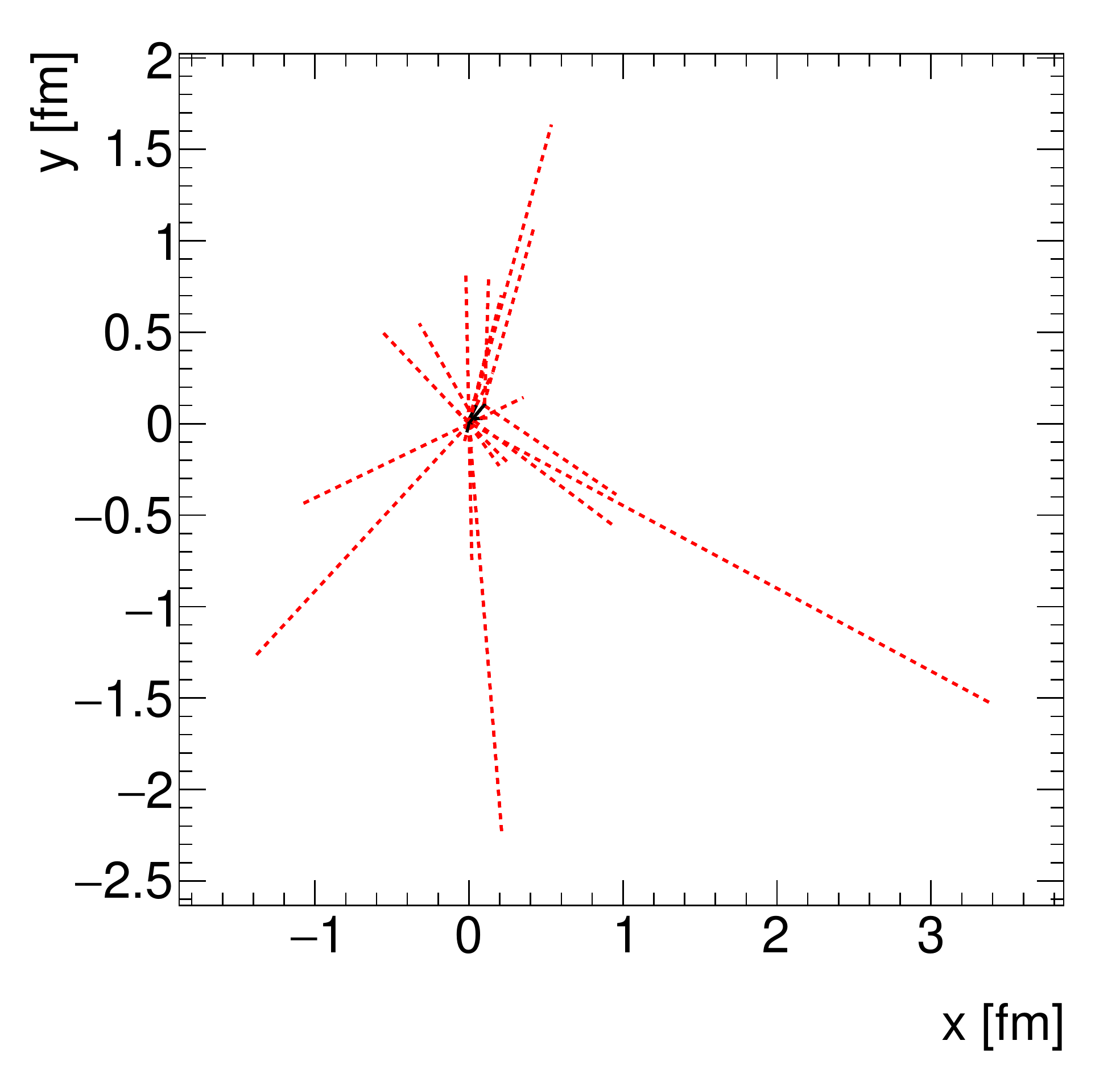}
\includegraphics[width=0.49\textwidth]{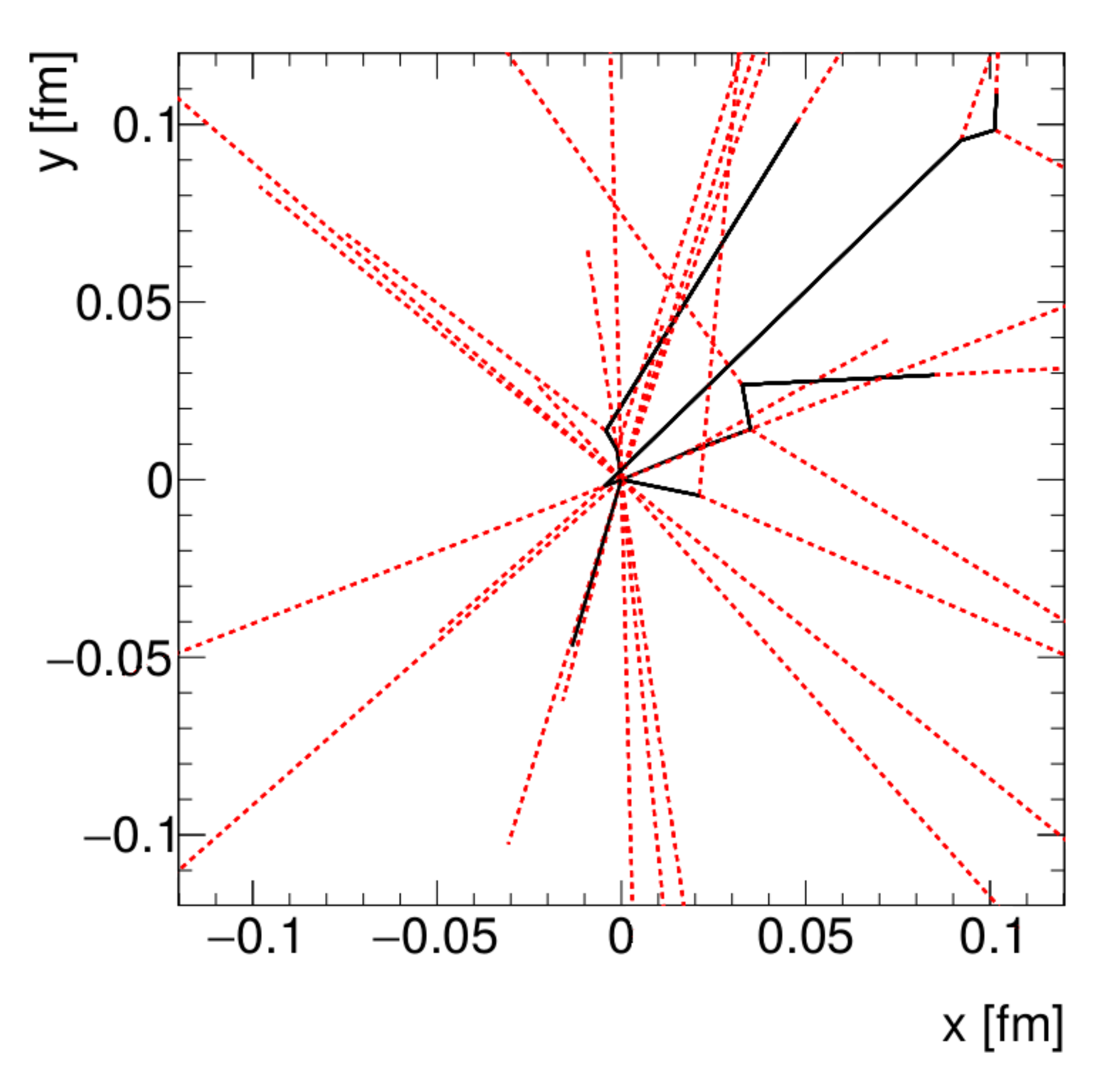}

\caption{
An example of a parton shower spacetime structure (i.e. neglecting spacetime structure of MPI) of a Minimum Bias event in the transverse plane generated with the minimum virtuality $\nu^2=1$~GeV$^2$.
The red dotted lines represent the evolution of the last particle in the parton shower while the
rest of the evolution is denoted by the black lines. Both panels show
the same event with the right panel magnifying the center of the event. 
}
\label{fig:Shower_tree}
\end{figure*}

\begin{figure*}[th]
\centering
\includegraphics[width=0.49\textwidth]{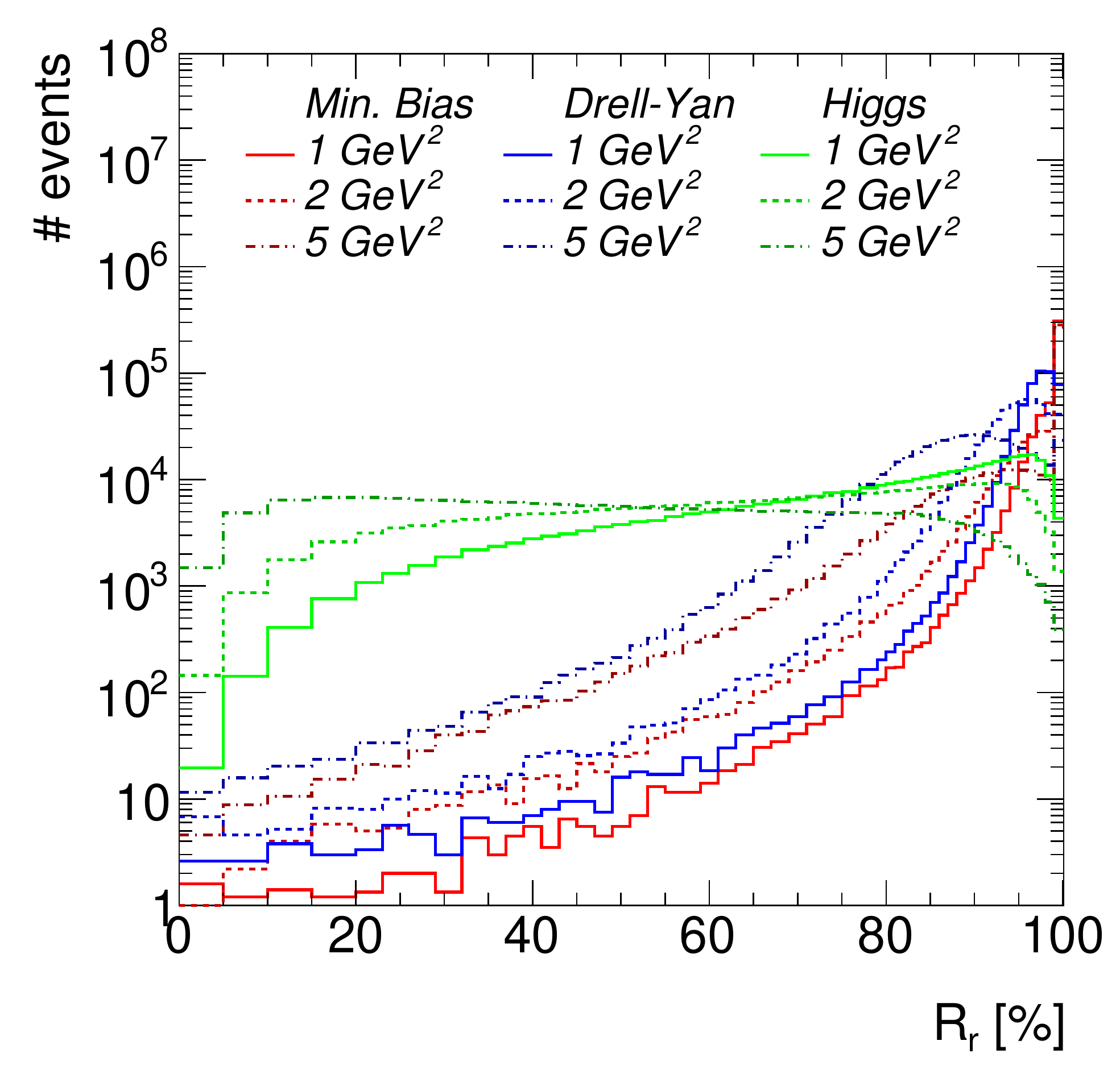}
\caption{
The ratio of the distance traveled by partons 
in the last step of evolution to the total distance (distance traveled during entire evolution).}
\label{fig:Percentages_r}
\end{figure*}

This is also visible in Fig.~\ref{fig:Shower_tree} where we show the
spacetime structure of a parton shower of a sample Minimum Bias event, with $\nu^2=1$~GeV$^2$,
neglecting the spacetime structure of the MPI positions. The final step 
distances are denoted by red dotted lines, while the intermediate steps are black solid lines.
In order to quantify this effect in Fig.~\ref{fig:Percentages_r} we show the 
ratio of distance traveled 
by partons in the last step of their evolution to the total distance 
(distance traveled during the entire evolution).
We see that in the case of both Minimum Bias and Drell-Yan processes for $\nu^2$ 
values similar to a typical parton-shower cutoff scale, 
i.e. below $2$ GeV$^2$, 90\% of the total distance is indeed due to 
the final step of the parton shower. 
In the case of the Higgs boson production, the distributions look very different. It is 
because in the simulation we took into account the decay lifetime of the Higgs boson, 
however when we neglect it, the distributions look very similar to
the two other processes.

To summarize, we can expect the fermi-scale parton shower 
and even further intermediate particle decay distances.
As such, these effects have to be included in spacetime colour reconnection model.
We also showed that tracing out the microscopic detail of 
the parton shower spacetime evolution is somewhat unnecessary, since only the 
low-energy scale of emissions (final steps) have any major impact on the spacetime position 
of partons, i.e. soft emissions close to the hadronization scale. 
Finally, it is important to stress that the Heisenberg uncertainty 
relations impose limits on how much simultaneous
energy–momentum and spacetime information one can have on an individual parton. 

These results should not be considered
as physical, but give us a benchmark of roughly what part of the event
simulation drives the creation of large separations in
distance between partons.

Instead, we propose a simpler model that assigns 
coordinates only to the very last partons of the parton shower, just before the hadronization.
This is in line with the uncertainty principle as the smearing 
is only visible for particles at a very 
soft scale. We may understand the partons' positions then as being smeared out around 
the scattering centres. This idea represents us taking the
semi-classical limit of the parton shower, and
generating coordinates in a similarly semi-classical manner.

\subsection{Parton shower coordinates}
As the partons propagate during the shower, we may assign a spacetime
propagation to their motion, but as we have shown above, 
these distances are only significant
at energy levels close to the hadronization scale.
As a consequence, we will only give spacetime coordinates to the partons
that remain at the end of the shower.
In our model of spacetime coordinates, we will not consider $z,t$ coordinates
and keep our discussion to the transverse plane.
We note that we have chosen the centre of mass frame
in order to construct our model, and to extend this to any given frame,
one need only transform the variables correspondingly.
All considerations below will be invariant to any boosts along the
$z$-axis.

Before the clusters are formed, each surviving parton from a 
given MPI scattering centre
receives an extra transverse propagation distance from the 
scattering centre coordinates. Instead of tracing out the positional
history of each parton during the shower, we take all partons at the
end of the shower and propagate them according to Eq. \ref{eq:decaylaw}. 
As argued above, this resembles a smearing of each partons' coordinate around
the scattering centre within its intrinsic uncertainty.

As discussed in Sec. \ref{sec:propagation}, at the end of the perturbative
shower, partons will have very small virtualities, meaning that using the
precise form of Eq. \ref{eq:spacetimePropagation} performs poorly. We instead
approximate the mean lifetime by considering the width term in the
denominator.
Each parton of species $p$ will automatically receive a
minimum virtuality, $\nu^2$, for their mean lifetime in their rest-frame:
\begin{equation}
    \tau_{0,p} = \frac{\hbar m_p}{\nu^2} .
    \label{eq:approxLifetime}
\end{equation}
This mean lifetime is derived from Eq.
\ref{eq:spacetimePropagation}, by taking the
on-mass shell limit - $\tau(q^2 = M^2) = \hbar/\Gamma$ and using the following form for the
width of the on-mass shell partons:
\begin{equation}
    \Gamma = \frac{\nu^2}{m_p} .
\end{equation}

With the mean lifetime from Eq. \ref{eq:approxLifetime}, we proceed as explained in Sec. \ref{sec:propagation},
using Eqs. \ref{eq:decaylaw} and \ref{eq:labCoords} to set each parton's
position relative to the MPI scattering centre that they originated from,
adding only the transverse coordinates of the propagation distance.

Eq. \ref{eq:approxLifetime} corresponds to a lab-frame mean lifetime of:
\begin{equation}
    \tau'_{0,p} = \gamma \tau_{0,p} = \frac{\hbar E_p}{\nu^2} ,
\end{equation}
where $E_p$ is the lab-frame energy of the given parton.
The main motivation for the mass dependence of the mean lifetime in Eq. (12) is that the decay distance of external light quarks is proportional to 
their energy (and independent of their mass) which is in agreement with expectations 
from the linear confining potential of QCD, see e.g. \cite{Bali:2000un} and
references therein, as well as other hadronization models such as 
the Lund string model \cite{Andersson:1983ia}.

As a result of this construction, quark-antiquark pairs produced during the
non-perturbative gluon splitting will receive the same spacetime position.
One may believe this leads to issues where colour 
reconnection wants to pair these partons together,
but \hw{} does not allow them to since they would be in a colour-octet
state \cite{Gieseke:2012ft, Reichelt:2017hts}. These partons will also have
slightly different rapidities, due to kinematics from the gluon splitting.

\begin{figure*}[th]
\centering
\includegraphics[width=0.49\textwidth]{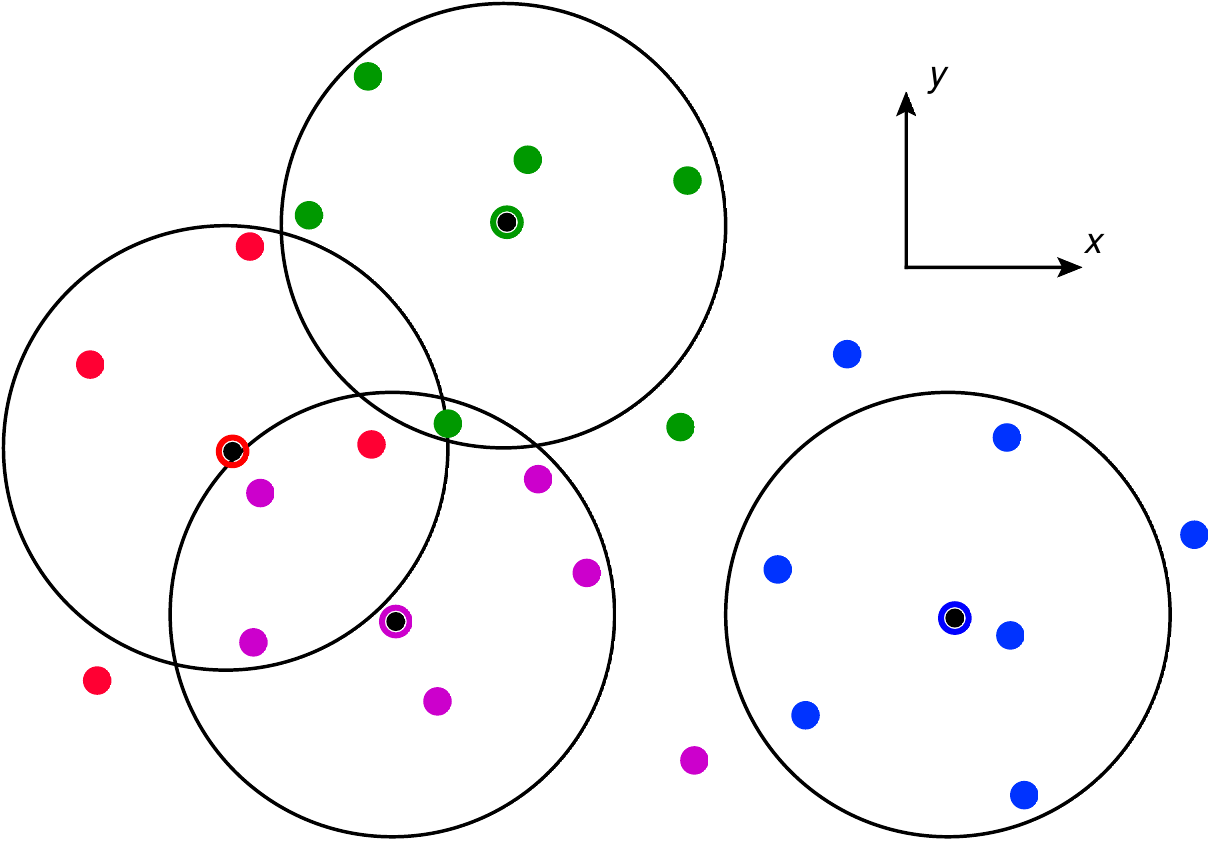}
\caption{A schematic diagram for how our model introduces 
transverse spacetime coordinates
for the multiple parton interactions (black points), and for the end of the
parton shower. Different coloured points
are partons from different, respectively ring-coloured MPI centres. The
thin black circles represent a characteristic scale for parton propagation about
the MPI centre.
}
\label{fig:stcrSummary}
\end{figure*}

Once all the partons have their new coordinates with respect to their 
MPI scattering centre, we then shift these coordinates using the
points produced from the MPI coordinate generator, as shown schematically 
in Fig. \ref{fig:stcrSummary}. The black points are the MPI centres, and
partons from those systems are spread by Eq. \ref{eq:labCoords}, 
around their respective centre.
Different coloured partons refer to partons originating from
different MPI systems.

\section{Spacetime Colour Reconnection}
With the transverse coordinates in place, we use this information
to perform and inform colour reconnection. We present the outline for plain spacetime
colour reconnection model, but we will use the baryonic spacetime model
for tuning and in the discussion in the rest of the paper.
\label{sec:spacetimeCR}

\subsection{Plain spacetime colour reconnection}
As mentioned in Sec. \ref{sec:cr}, the measure for allowing plain
colour reconnection is the sum of invariant cluster masses before and after,
and the reconnection is given by a flat tuned weight.
However, there is at least one major issue with this construction:
this measure aims to reconnect cluster
constituents so that they are closer in momentum space, but without any
input from spacetime which would perhaps prohibit a causally-disconnected
colour reconnection.

Using the coordinates we have introduced in Sec. \ref{sec:coords},
we now define the following spacetime-inspired measure for a single cluster
with constituents $i,j$:
\begin{equation}
    R_{ij}^2 = \frac{\Delta r_{ij}^2}{d_0^2} + \Delta y_{ij}^2 ,
    \label{eq:spacetimeMeasure}
\end{equation}
where $d_0$ is the characteristic length scale for colour reconnection
in our spacetime model, which is a tunable parameter.
$\Delta r_{ij}^2 = (\vec{x}_{\perp,i} - \vec{x}_{\perp,j})^2 $ 
is the transverse spacetime separation squared between the constituent
quarks.
We include rapidity differences in Eq. \ref{eq:spacetimeMeasure}. This
is inspired by conventional jet algorithms, where we replace the azimuthal
separation $\Delta \phi_{ij}^2$ with transverse separation.
The parameter $d_0$ effectively acts as a measure to increase the 
importance of transverse to longitudinal components.
The measure in Eq. \ref{eq:spacetimeMeasure} captures the
transverse separation between the constituents
and their longitudinal separation.

Using the measure from Eq. \ref{eq:spacetimeMeasure}, we proceed in the same
fashion as Eq. \ref{eq:massMin}, by minimizing the sum of the pairing of
cluster constituents. 
For a given cluster, we pick the candidate cluster that minimizes the 
measure the most. If the sum of the cluster separations is smaller
after a possible reconnection:
\begin{equation}
    R_{q\qb'} + R_{q'\qb} < R_{q\qb} + R_{q'\qb'} ,
    \label{eq:recoCriterion}
\end{equation}
then we accept the reconnection
with a flat probability, $p_{\mathrm{M, reco}}$.
A similar model
was studied earlier in \cite{Rohrdiploma}.

\subsection{Baryonic spacetime colour reconnection}
Baryonic spacetime colour reconnection uses the algorithm from
\cite{Gieseke:2017clv}, and outlined in Sec.
\ref{sec:cr}. The partners for mesonic and baryonic colour reconnection
are found by using the projection onto a given cluster's quark axis.

If instead we find a baryonic reconnection, we cannot directly compare the
sum of Eq. \ref{eq:spacetimeMeasure} for the constituents of the clusters
before and after colour reconnection, since we would be starting with 
3 clusters - each with 2 partons - and ending with 2 clusters with 3 partons,
and the distance measure is an ill-defined quantity in the latter situation.

In the ordinary baryonic colour reconnection algorithm, 3-component
clusters, once formed, are reduced to a quark-diquark system, 
where the diquark system is chosen as the pair of quarks with the
lowest total invariant mass. In keeping with our spacetime
paradigm, we choose the pair as the closest in spacetime. 
Given 3 mesonic clusters, we look
at the set of triplets $\{q_1,q_2,q_3\}$ and select the pair that are
closest - calculated via Eq. \ref{eq:spacetimeMeasure}, and similarly for 
the set of antitriplets. We choose these partons to become a diquark system,
with their constituents' mean spacetime position and rapidity. 

We allow baryonic reconnection if the following criterion is true:
\begin{equation}
    R_{q,qq} + R_{\bar{q},\bar{q}\bar{q}} < R_{q,\bar{q}} + R_{qq,\bar{q}\bar{q}} ,
\end{equation}
which is analogous to Eq. \ref{eq:recoCriterion}, and 
we accept this reconnection with probability $p_{\mathrm{B,reco}} = w_b$.
If the reconnection is rejected, all three candidate clusters
remain ordinary mesonic clusters.

We note that the baryonic spacetime colour reconnection has a bias
for using rapidity as its first discriminating factor when searching for
potential partners. However, we hope that, 
by using the extra information provided
by the transverse separation between constituents, we will be able 
to improve upon the
original baryonic colour reconnection model, 
especially in larger systems like heavy ion collisions.

\begin{figure*}
\centering
\includegraphics[width=0.82\textwidth]{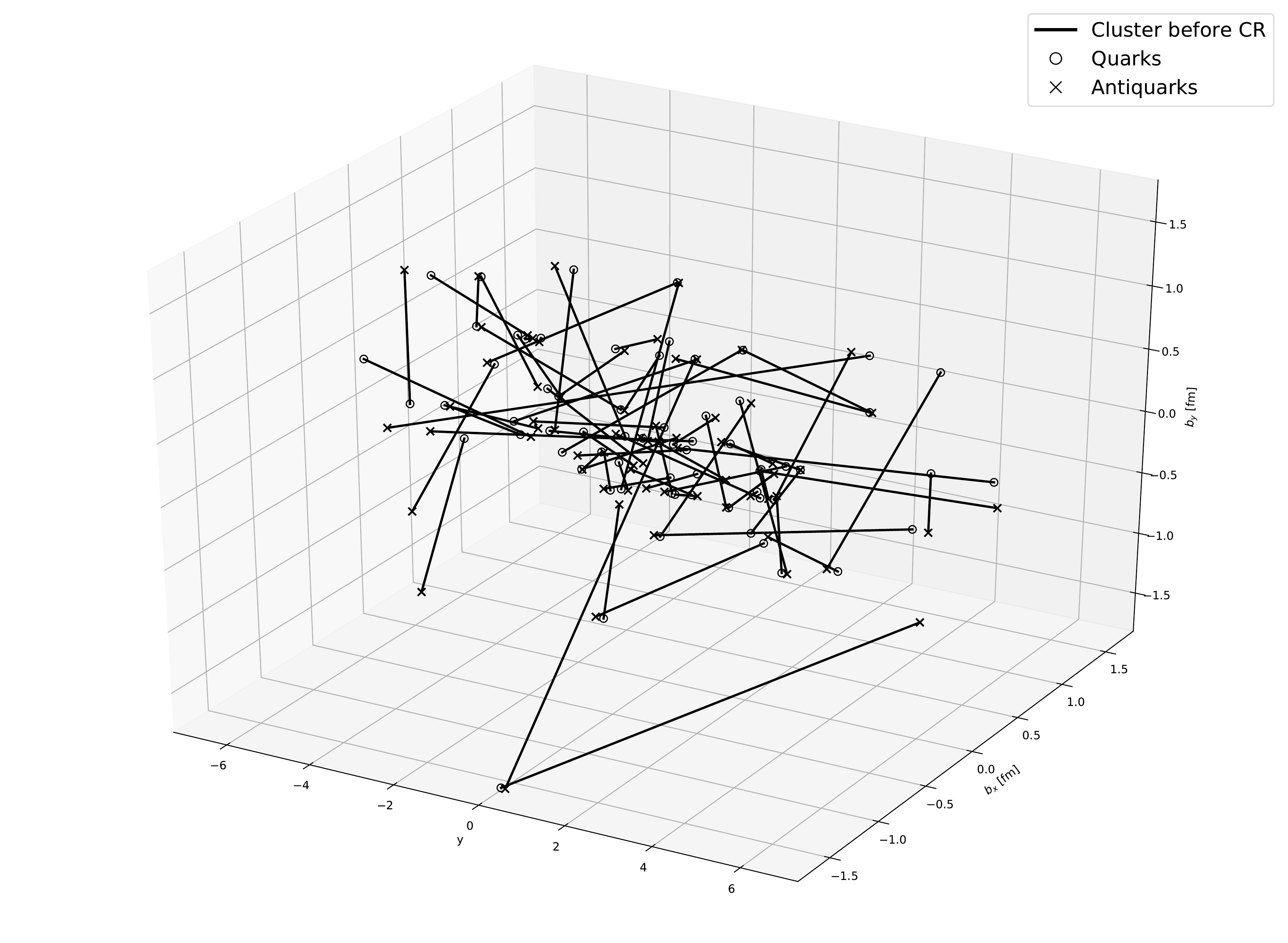}\\
\includegraphics[width=0.82\textwidth]{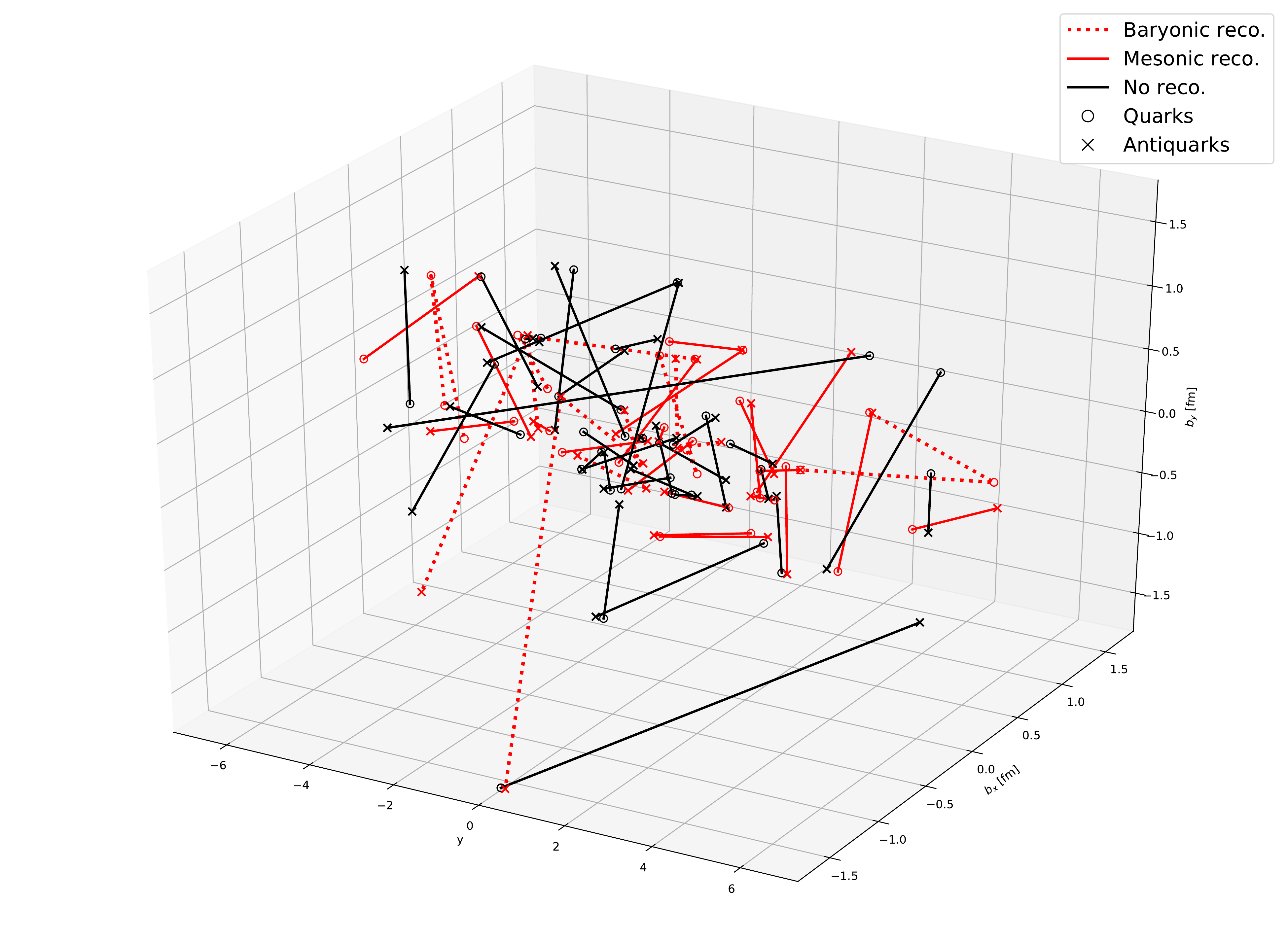}
\caption{The colour-topology of a sample Minimum Bias 
event in rapidity and 
transverse spacetime coordinates, before (top) and after (bottom)
colour reconnection.
The parameters used for reconnection here are $\nu^2 = 1$ GeV$^2$,
$d_0 = 0.5$ fm, and baryonic reconnection weight $w_b = 0.5$.
Black lines correspond to clusters which are automatically produced from
the parton shower and which have not undergone any
colour reconnection, while red lines are the newly rearranged 
(dotted lines) baryonic and (solid lines) mesonic clusters.
}
\label{fig:colourReconnection}
\end{figure*}

To see the spacetime picture of an event, we have produced Fig.
\ref{fig:colourReconnection}, which highlights the spacetime coordinate
generation procedure outlined in Sec. \ref{sec:coords}.
In the upper panel of Fig. \ref{fig:colourReconnection}, we have plotted all the clusters formed 
from the non-perturbative gluon splitting at the end of the shower, 
before any
colour reconnection. The points in the plots represent cluster constituents,
and the connecting lines represent the clusters.

Performing baryonic spacetime colour reconnection, using $\nu^2 = 1$ GeV$^2$,
$d_0 = 0.5$ fm, and $w_b = 0.5$, on this event then produces the lower panel in Fig. \ref{fig:colourReconnection},
where we have highlighted the different types of clusters. Red lines correspond
to rearranged clusters: (dotted) baryonic, and (solid) mesonic, while black
lines are untouched clusters.

\section{Modifications to the Existing Model}
\label{sec:modifications}

While incorporating spacetime coordinates into the \linebreak
\hw{} MPI model, we have had to modify parts of the
original implementation. These changes are of a more
general nature than the specifics of our model.
As we wish to focus on the changes that our model has, we
will report the changes in a separate 
contribution \cite{Bellm:2019icn}.
We summarize the most relevant modifications below:
\begin{itemize}
    \item The kinematics is improved and produces the wanted inclusive spectrum.
    \item Introduction of diffraction ratio $R_{\mathrm{Diff}}$ parameter for better tuning performance.
    \item Cross-section handling takes into account the diffractive cross section to calculate the eikonalised cross sections.
    \item The dummy process used by \hw{} in Minimum Bias 
    events is replaced to contain only initial state quarks.
    \item The partner finding process and scale setting are modified with respect to the standard \hw{} mode.  
\end{itemize}
The effects of these changes and their discussion are postponed to \cite{Bellm:2019icn}.

\section{Tuning}
\label{sec:tuning}

\begin{figure*}[th]
\centering
\includegraphics[width=0.32\textwidth]{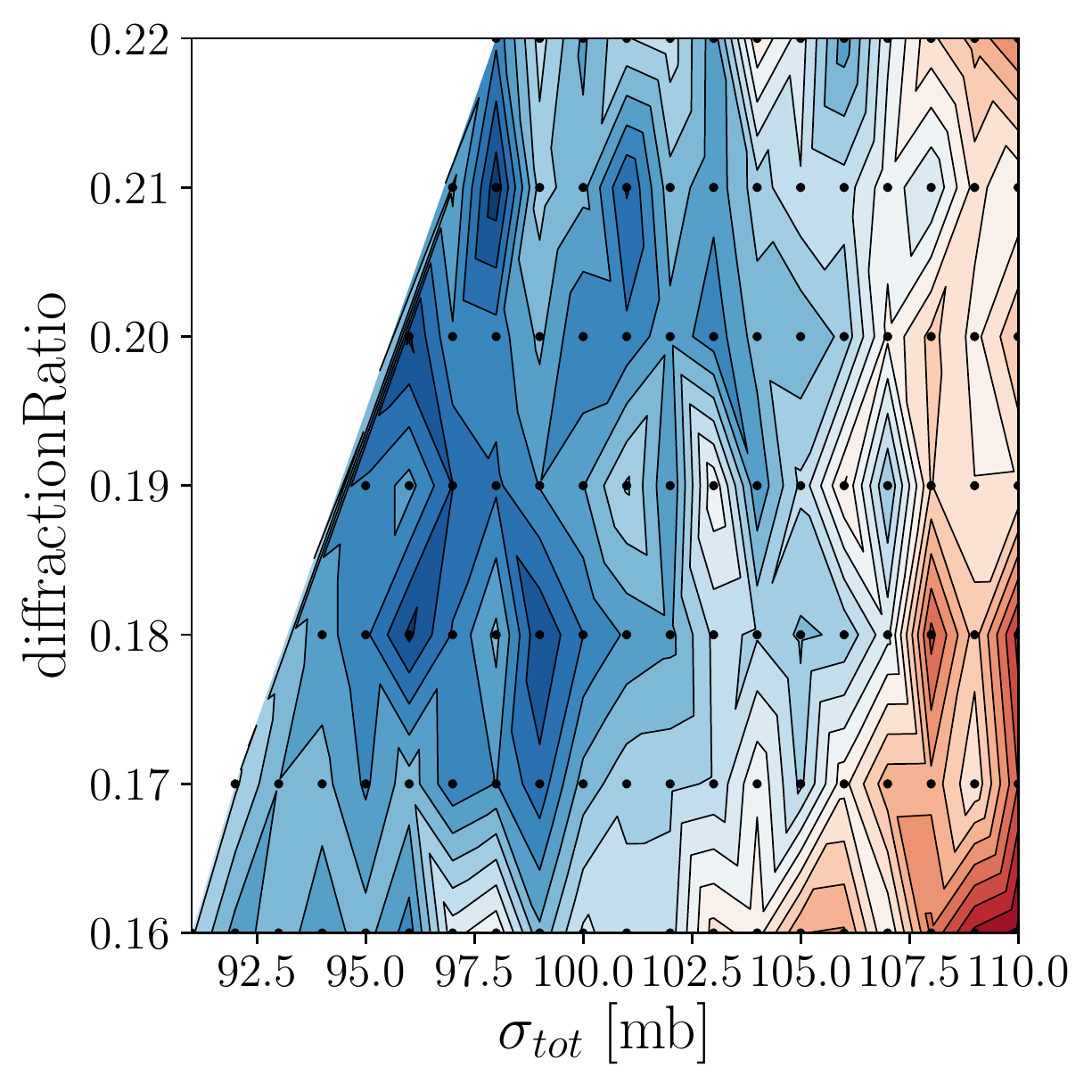}
\includegraphics[width=0.32\textwidth]{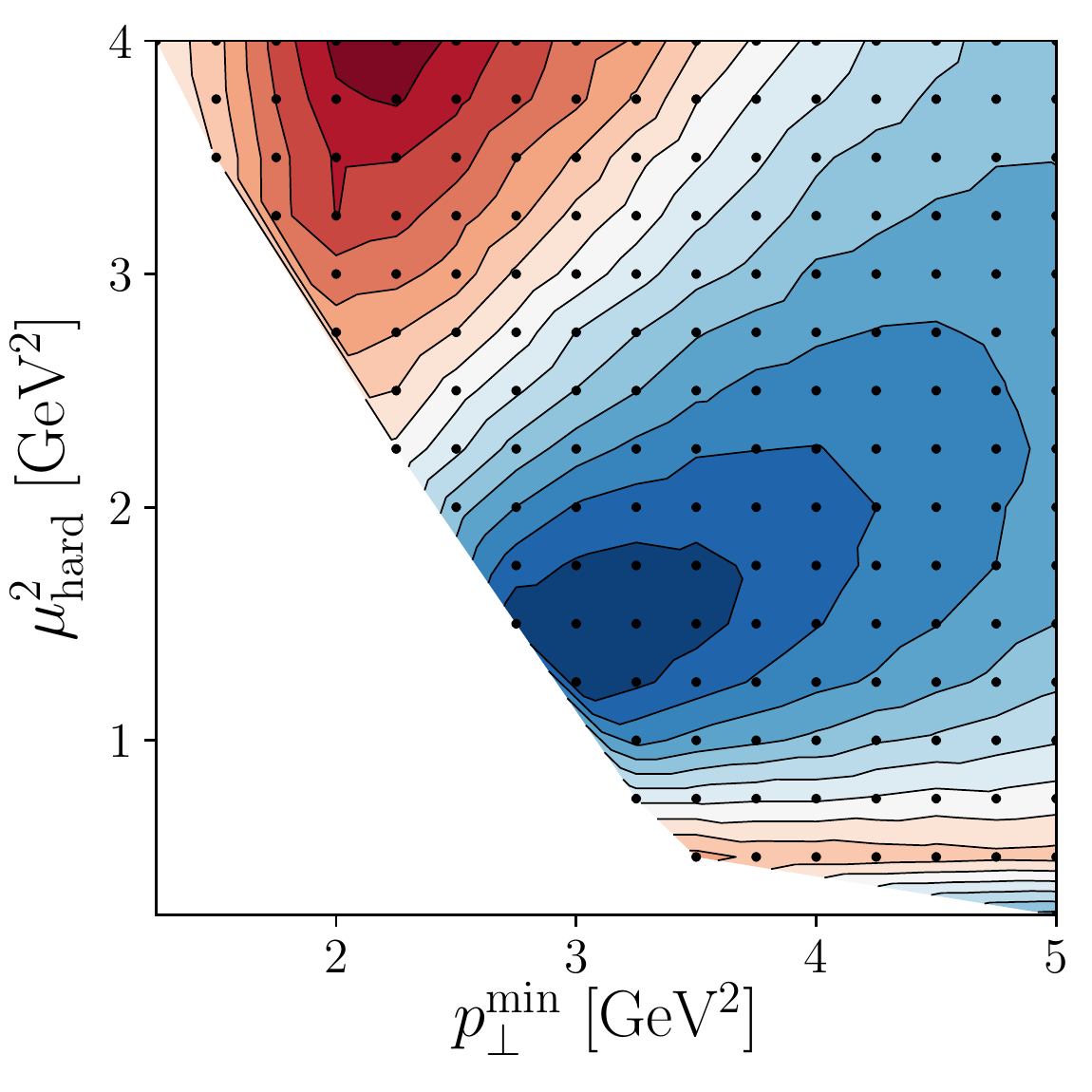}
\includegraphics[width=0.32\textwidth]{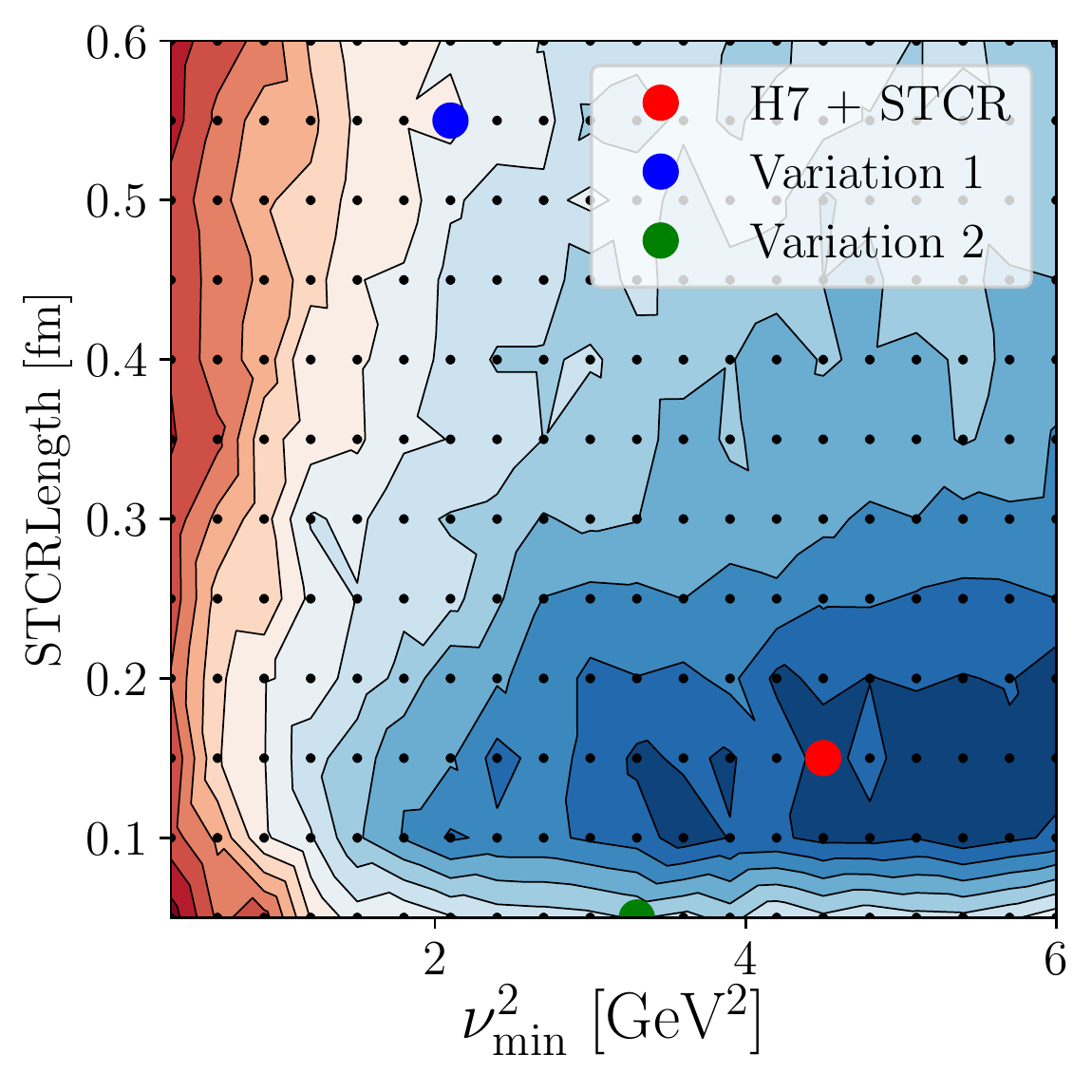}
\caption{$\chi^2$-planes for parameter sets ($R_{\mathrm{Diff}}$, $\sigma_{\mathrm{tot}}$), 
($\mu^2_{\mathrm{hard}}$, $p_{\perp}^{\mathrm{min}}$) and ($\nu^2$, $d_0$).
Bluer areas in the $\chi^2$ contour plots correspond to smaller $\chi^2$ values.
In the right plot we pick three parameter pairs to define variations 
to be shown in the data comparison, see Figs. \ref{fig:Fig1} to \ref{fig:Fig7}.}
\label{fig:2dplanes}
\end{figure*}

We started the tuning process within the Autotunes \cite{Bellm:2019owc}
framework that internally makes use of the Rivet and 
Professor frameworks \cite{Buckley:2010ar,Buckley:2009bj} for 
Monte Carlo event generators. 
To elucidate the effects of parameter variations, we illustrate 
the modifications in $\chi^2$-values in Fig. \ref{fig:2dplanes}. 
Here, we show by variation of strongly correlated parameter pairs 
where the minimum of the parameters are located. 
The white spaces in the planes for the parameter sets 
($R_{\mathrm{Diff}}$, $\sigma_{\mathrm{tot}}$) and 
($\mu^2_{\mathrm{hard}}$, $p_{\perp}^{\mathrm{min}}$) are regions 
in parameter space where the model fails to fit the soft and hard
cross-sections without violating the total cross-section. 
In the left $\chi^2$-plane, we added lines to mark the total cross 
sections that are predicted by the Donnachie and Landshoff
model, where 
DLMode 1 refers to \cite{Donnachie:1992ny}, DLMode2 refers 
to \cite{Donnachie:1992ny} but normalized to \cite{Abe:1993xy}.
\footnote{A third mode that is implemented in Herwig 7 that would 
refer to \cite{Donnachie:2004pi} would predict a total cross section 
of $\sigma_{\mathrm{tot}}=120.496~\mathrm{mb}$ and is not acceptable
with our tuning.} 

In the ($\nu^2$, $d_0$)-plane, we define 
three parameter points to be used in the later
data comparisons. 
The red point, corresponding to the best fit value 
($\nu^2=4.5$ GeV$^2$, $d_0=0.15$ fm) 
will be referred to as ``H7 + STCR".
To show variations in the spacetime model, we choose two other points:
blue - ($\nu^2=2.1$ GeV$^2$, $d_0=0.55$ fm),
and green - ($\nu^2=3.3$ GeV$^2$, $d_0=0.05$ fm).
These two points will be referred to as ``Variation 1" 
and ``Variation 2" in the following. 

We compared the model in the tuning procedure to data from 
\cite{Adam:2015qaa,Abelev:2012sea,Aad:2010ac,Aad:2010fh,Khachatryan:2015gka} 
and the red parameter point in Fig. \ref{fig:2dplanes}
corresponds to the parameters that are reflected in Tab. \ref{tab:tune}.
\begin{table}
\centering
\begin{tabular}{lcccccc}
      $\sigma_{\mathrm{tot}}~[\mathrm{mb}]$ & $R_{\mathrm{Diff}}$&
      $p^{\mathrm{min}}_{\perp}~[\mathrm{GeV}]$&$\mu^2_{\mathrm{hard}}~[\mathrm{GeV}^2]$\\
\midrule
 $96.0$ & $0.2$  & $3.0$  & $1.5$\\
 \\
 $\nu^2~[\mathrm{GeV}^2]$ & $d_0~[\mathrm{fm}]$ & $w_b$ &  ($\mu^2_{\mathrm{soft}}~[GeV^2]$)  \\
 \midrule
 $4.5$ & $0.15$ & $0.98$ & $0.254$ \\
\end{tabular}
  \caption{The newly tuned parameters for Minimum Bias 
  simulation and our
  baryonic spacetime colour reconnection model. The top row is the
  re-tuned parameters of the old \hw{} Minimum Bias model. 
  The bottom row is the three new parameters of the spacetime
  components of our model, and a determined parameter of the old
  model.}
  \label{tab:tune}
\end{table}
The parameters in the first row have been previously included
in the \hw{} Minimum Bias model. $R_{\mathrm{Diff}}$ was not explicitly part of the regular 
model in \hw{} but was effectively tuned as the amplitude of 
the non-diffractive cross section. $p^{\mathrm{min}}_{\perp}$ is the 
cut on the transverse momentum where the hard MPI component, 
described by perturbative QCD $2\to 2$ process is taken over by 
the soft, multi-peripheral MPI model \cite{Bahr:2009ek,Gieseke:2016fpz}.
The parameter for the inverse proton radius is $\mu^2_{\mathrm{hard}}$ 
and is communicated together with the determined (not tuned) parameter 
for the soft inverse radius $\mu^2_{\mathrm{soft}}$ to the 
MPI coordinate generator.

The parameters in the second row are the three new parameters introduced
for our spacetime model. First, the minimum virtuality $\nu^2$, which 
dictates the traveling of the final 
partons after the shower step, takes a rather large value $4.5$ GeV$^2$
in comparison to the parton shower $Q^2$ cutoff.

Second, the colour reconnection distance scale $d_0$ in 
Eq. \ref{eq:spacetimeMeasure} has a tuned value of 0.15 fm.
This length scale is the strength of 
the transverse component of the spacetime measure
relative to the rapidity component. It can also be considered the
characteristic length scale of colour reconnection in the 
transverse plane in our model.

Finally, the baryonic colour reconnection probability weight
$w_b$, after tuning, has a value of 0.98.
This seems to be very large but the model, 
as described in \cite{Gieseke:2017clv},
already makes strong restrictions on the possible cluster configurations 
such that the cluster triplets that are potential candidates for the
baryonic reconnection are strongly favoured. 

We have kept the probability for strangeness production during the
non-perturbative gluon splitting as the tuned value from
\cite{Gieseke:2017clv}, although there have been recent developments
in the description of non-perturbative \linebreak
strangeness production in cluster
hadronization \cite{Duncan:2018gfk}. We leave a full retune of all the
hadronization parameters to future work.

\section{Results}
\label{sec:results}

\begin{figure*}[th]
\centering
\includegraphics[width=0.49\textwidth]{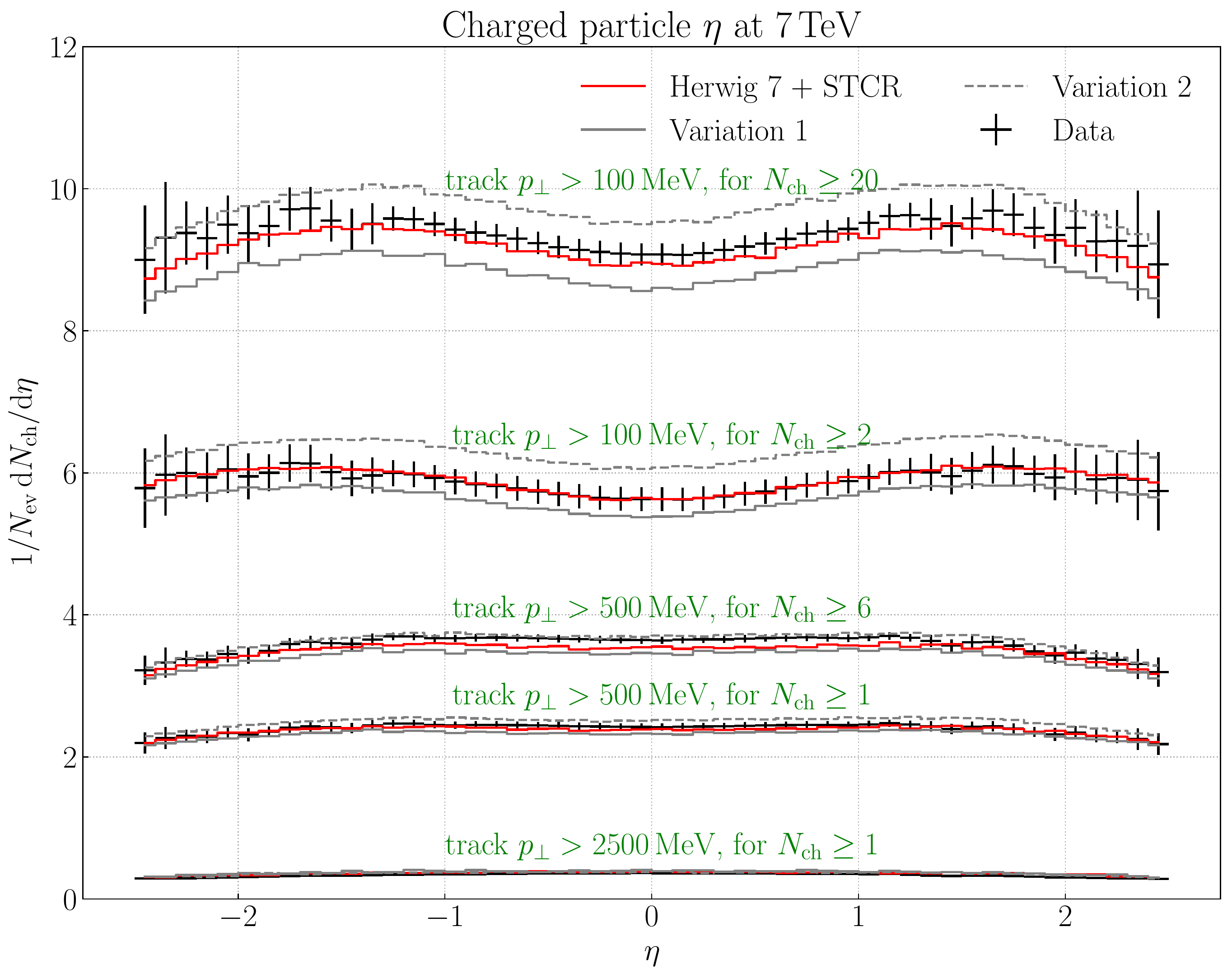}
\includegraphics[width=0.49\textwidth]{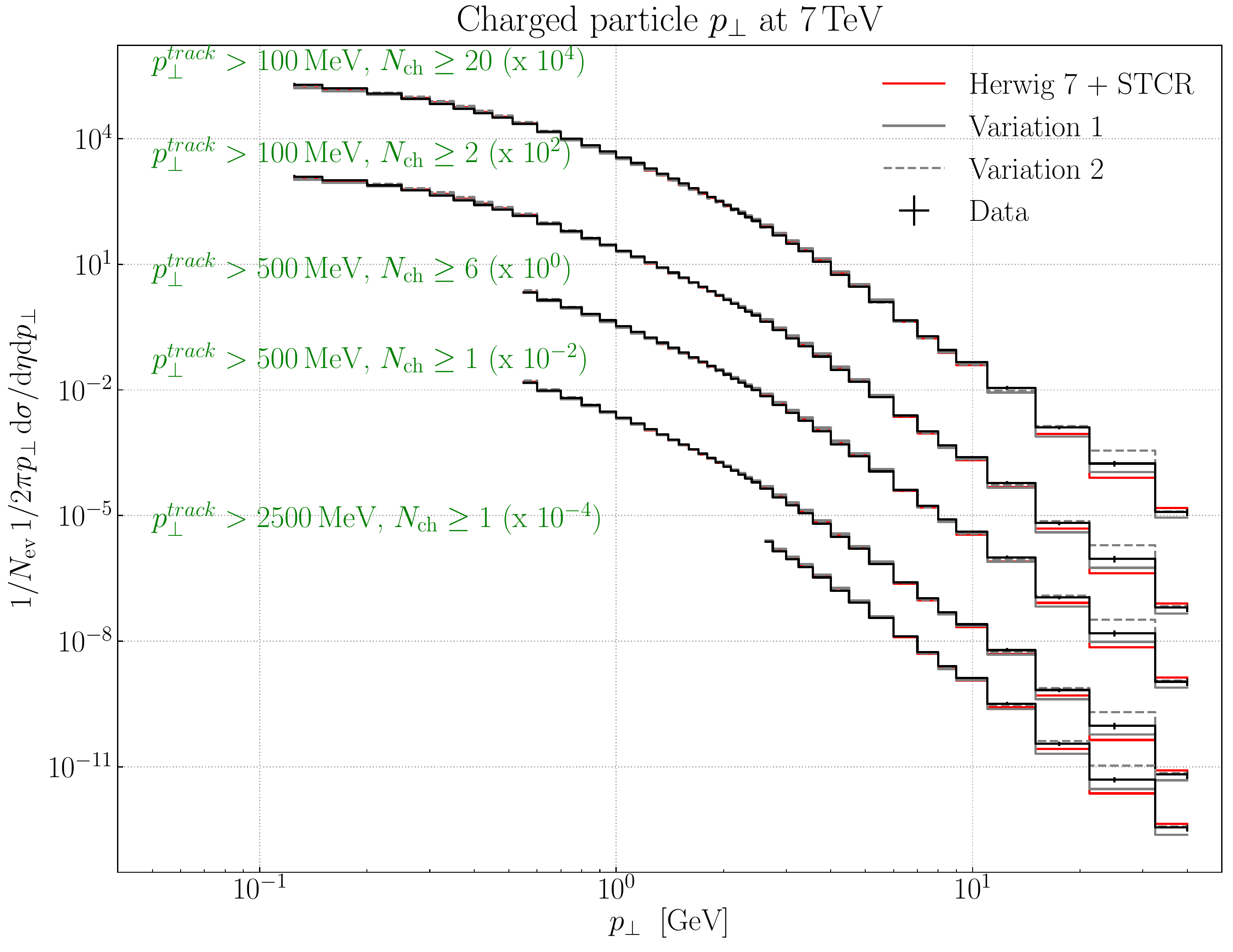}
\caption{Charged particle spectrum against rapidity and transverse
momentum for various leading track $p_{\perp}$ and number 
of charged particle $N_{\mathrm{ch}}$ slices. An overall good agreement with 
data is found. The variation is purely 
in the spacetime length and minimum virtuality parameters of our model
as defined in Fig. \ref{fig:2dplanes} and in the corresponding text. }
\label{fig:Fig1}
\end{figure*}

In this section, we describe the data comparison of the tuned parameter set. 
In Fig. \ref{fig:Fig1}, we have collated various cuts on the
track momentum, and similarly on the minimum number of charged
particles 
for the rapidity and transverse momentum distributions as measured
in \cite{Aad:2010ac}. Beside the central parameter set (red), 
we also show the results of the variations as gray lines 
(solid and dashed). 
These are crucial observables for the description of Minimum Bias
and soft physics, and we find that the model is perfectly capable 
at describing the distributions. 

\begin{figure*}[th]
\centering
\includegraphics[width=0.99\textwidth]{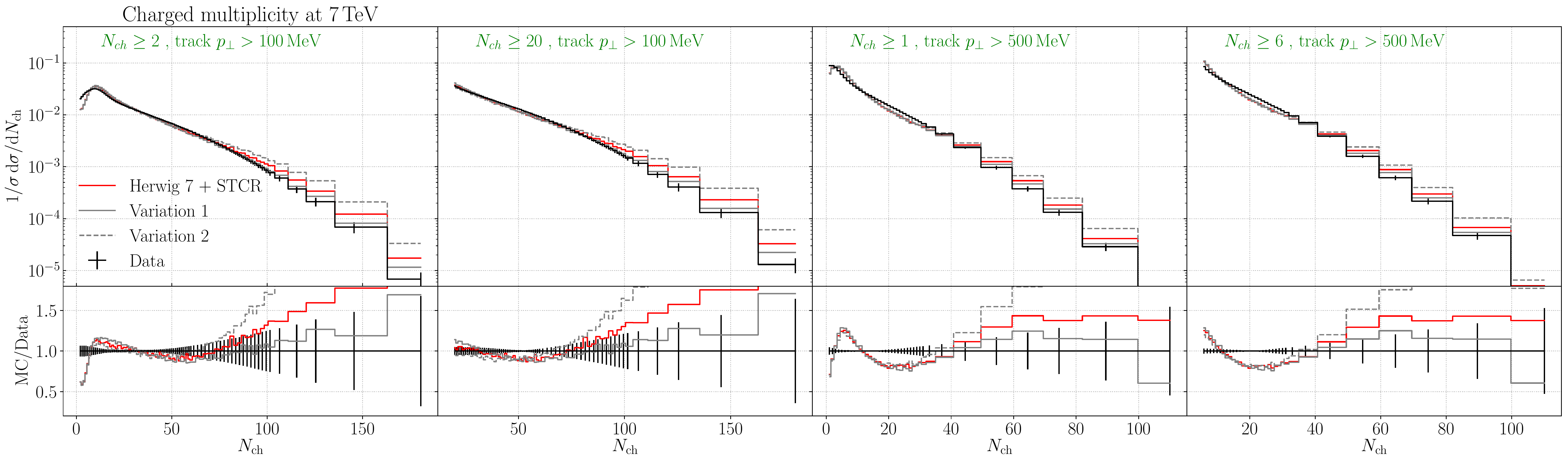}
\caption{Differential cross-section with respect to the number of charged particles as measured by \cite{Aad:2010ac}.}
\label{fig:Fig3}
\end{figure*}

In Fig. \ref{fig:Fig3}, we compare the differential cross-section with respect to the 
number of charged particles as measured by \cite{Aad:2010ac} with our model's results. 
We observe that for high charged particle multiplicity the central line 
overshoots the data and that ``Variation 1" is closer to the central data line. 
With the increased $d_0$ in ``Variation 1", the colour reconnection probability is increased. 
For a high number of additional scatters, the probability is increased to produce smaller clusters 
and therefore less particle production in the cluster fission and decay processes.

\begin{figure*}[th]
\centering
\includegraphics[width=0.49\textwidth]{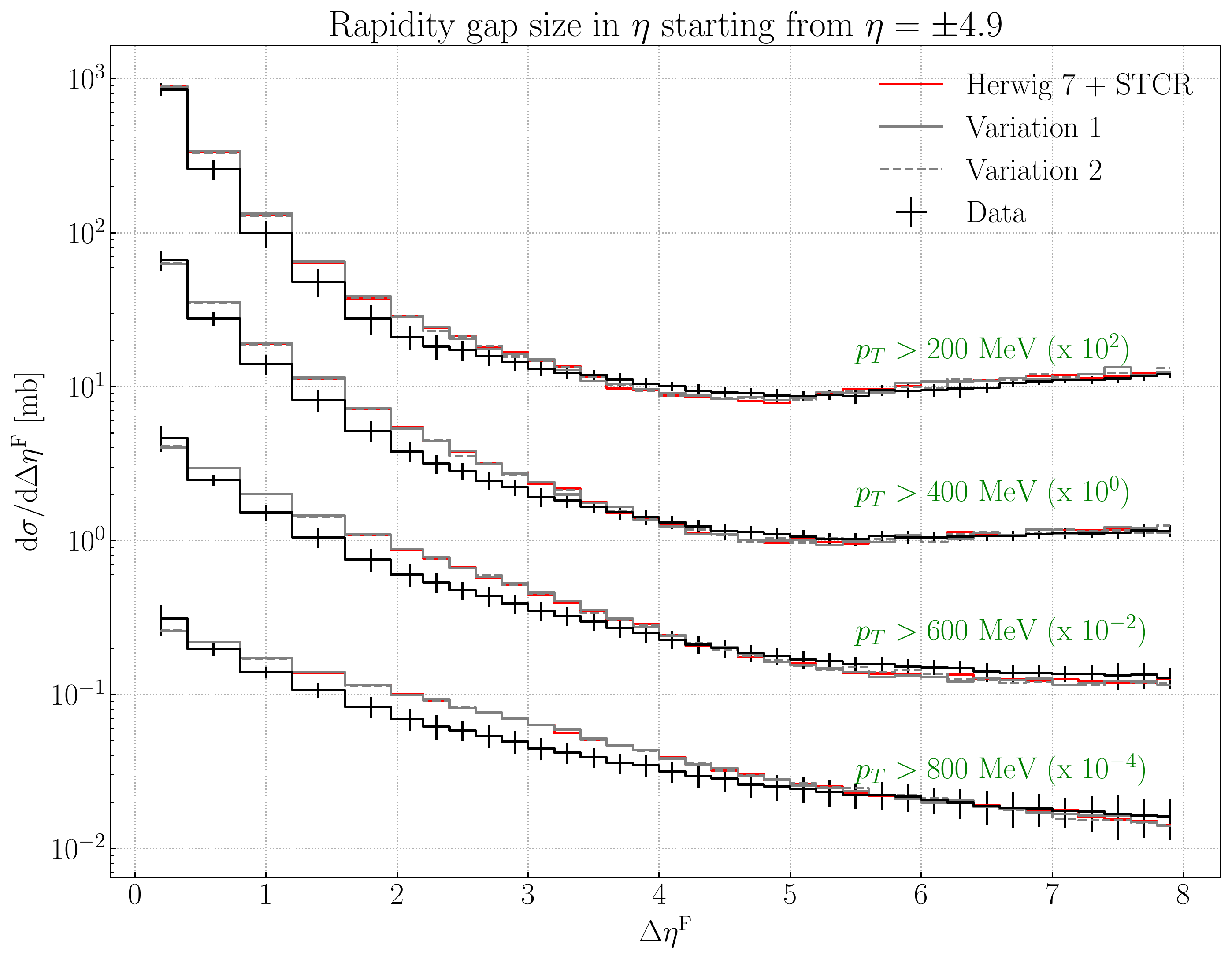}
\includegraphics[width=0.49\textwidth]{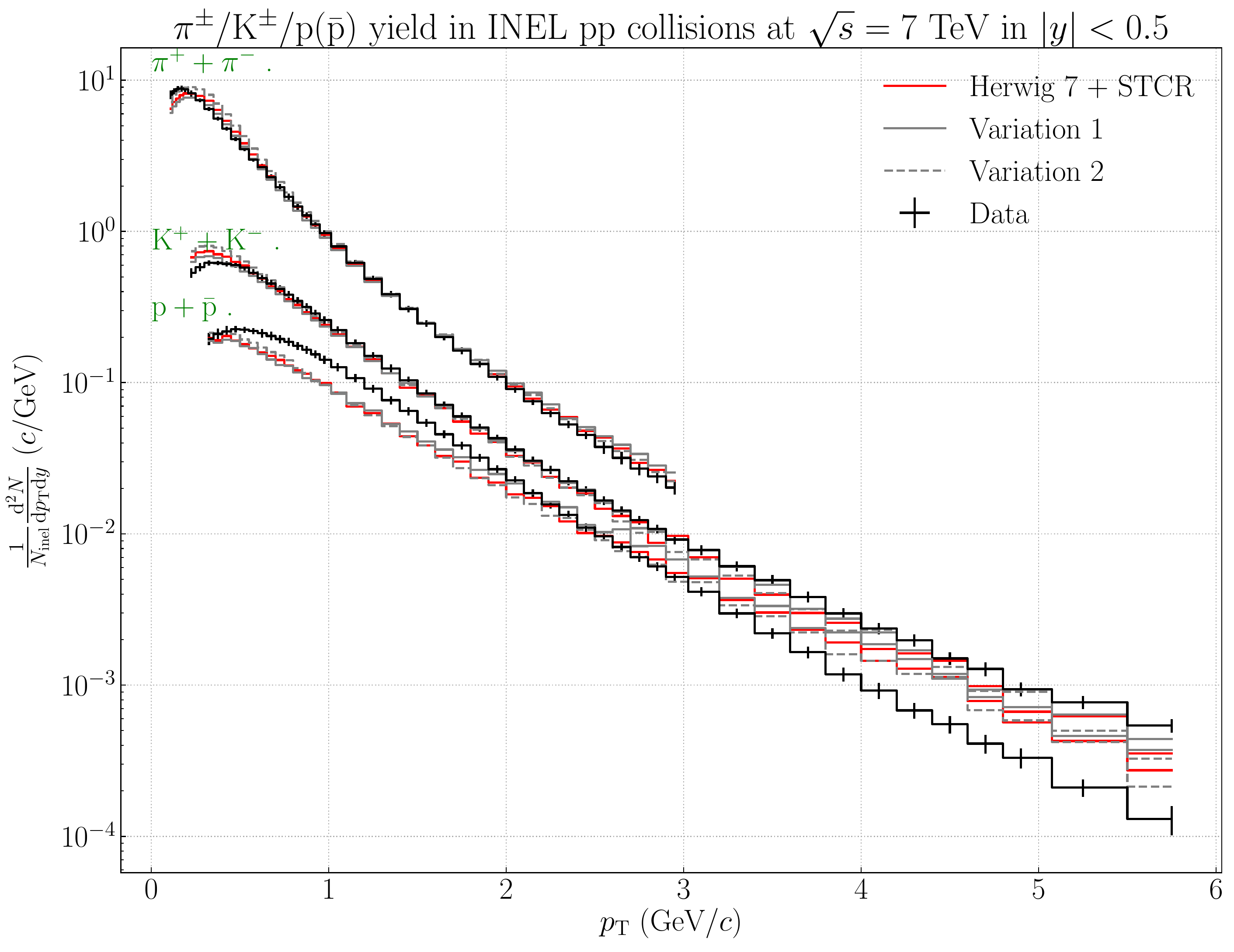}
\caption{Predictions for the rapidity gap fraction and the pion, kaon, and proton yields as measured by \cite{Aad:2012pw} and \cite{Adam:2015qaa}.
Variations in the spacetime components of the model show very little impact on the results.}
\label{fig:Fig4}
\end{figure*}

To illustrate examples of observables that are hardly modified by 
the variations in the spacetime 
components of the model, we show in Fig. \ref{fig:Fig4}
the measured rapidity gap fraction and
the pion, kaon, and proton yields as measured by \cite{Aad:2012pw} and \cite{Adam:2015qaa}. Variations in the spacetime components 
of the model have very little impact on these observables.
The rapidity gap for small values is mostly driven by the 
hard and soft MPI that could 
potentially be modified but is known to be relatively invariant
to colour reconnection effects. 
The tail of the rapidity gap cross section is mainly filled by double
and single diffraction, which are not modified by the smearing 
of the MPI collision centers. 
The fairly poorly described proton yield will be the subject of 
further studies. 

Typical observables that are used to verify the description of MPI
models in underlying event measurements are 
the angle of the particle production with respect to the 
leading track as well as the average sum of transverse momenta in the region towards, away,
and transverse to the leading track. 
Comparing our model to data measured at the ATLAS collaboration
\cite{Aad:2010fh}, we find that
the turn on behaviour, $p_{\perp} < 2.5 $ GeV for the leading track, is slightly too low. This has also been seen in the previous Herwig models.
For leading tracks above $2.5$~GeV, the average transverse momentum sum is about 10\% too large. 
This can also be seen in the radial dependence with respect to
the leading track.
In the Herwig MPI model, there is no azimuthal correlation between
the additional scatters. Herwig's only mechanism to correlate the
additional scatters is the colour reconnection. 
Introducing methods to correlate these scatters, as well as 
correlate them angularly, is left to future work.

\begin{figure*}[th]
\centering
\includegraphics[width=0.49\textwidth]{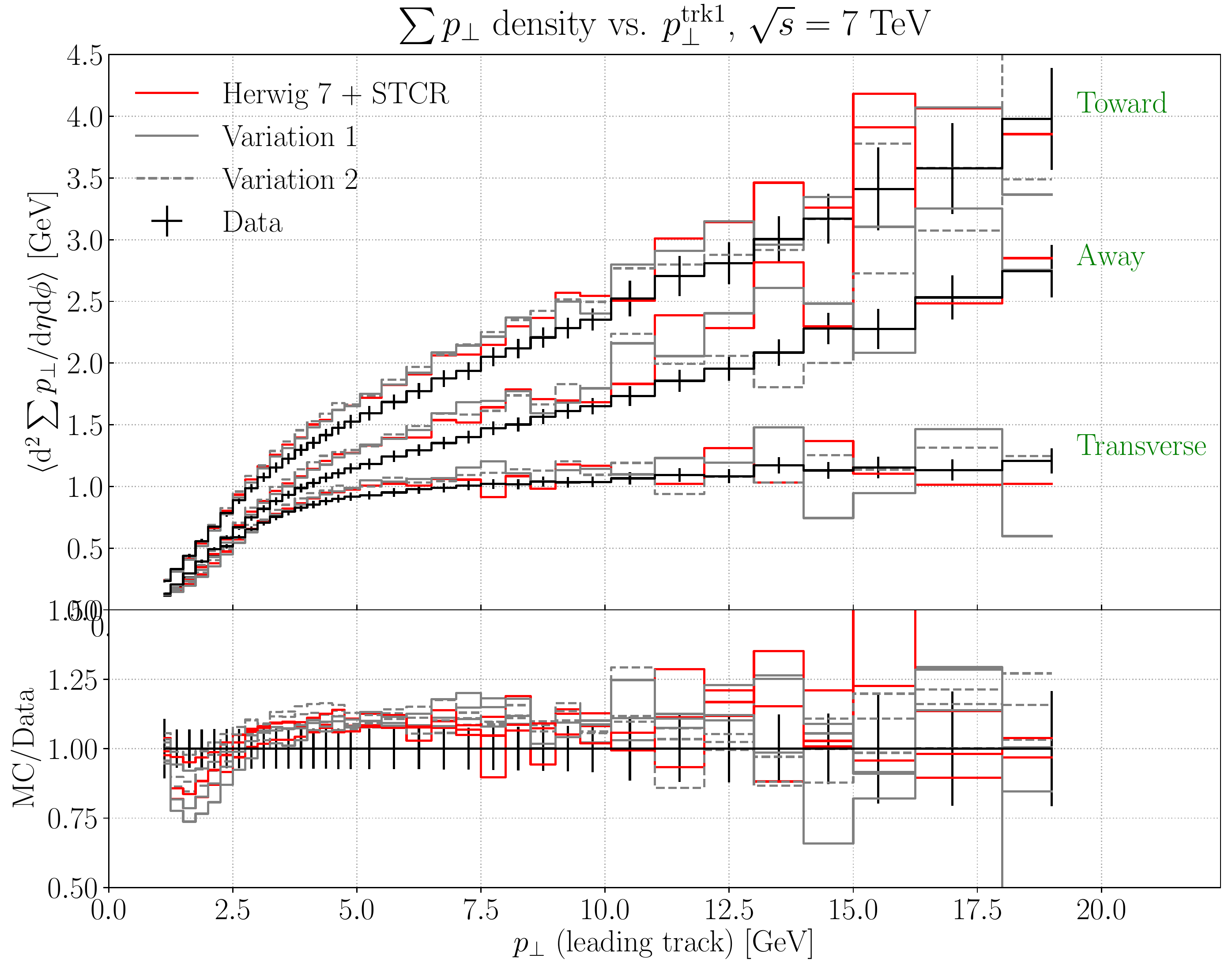}
\includegraphics[width=0.49\textwidth]{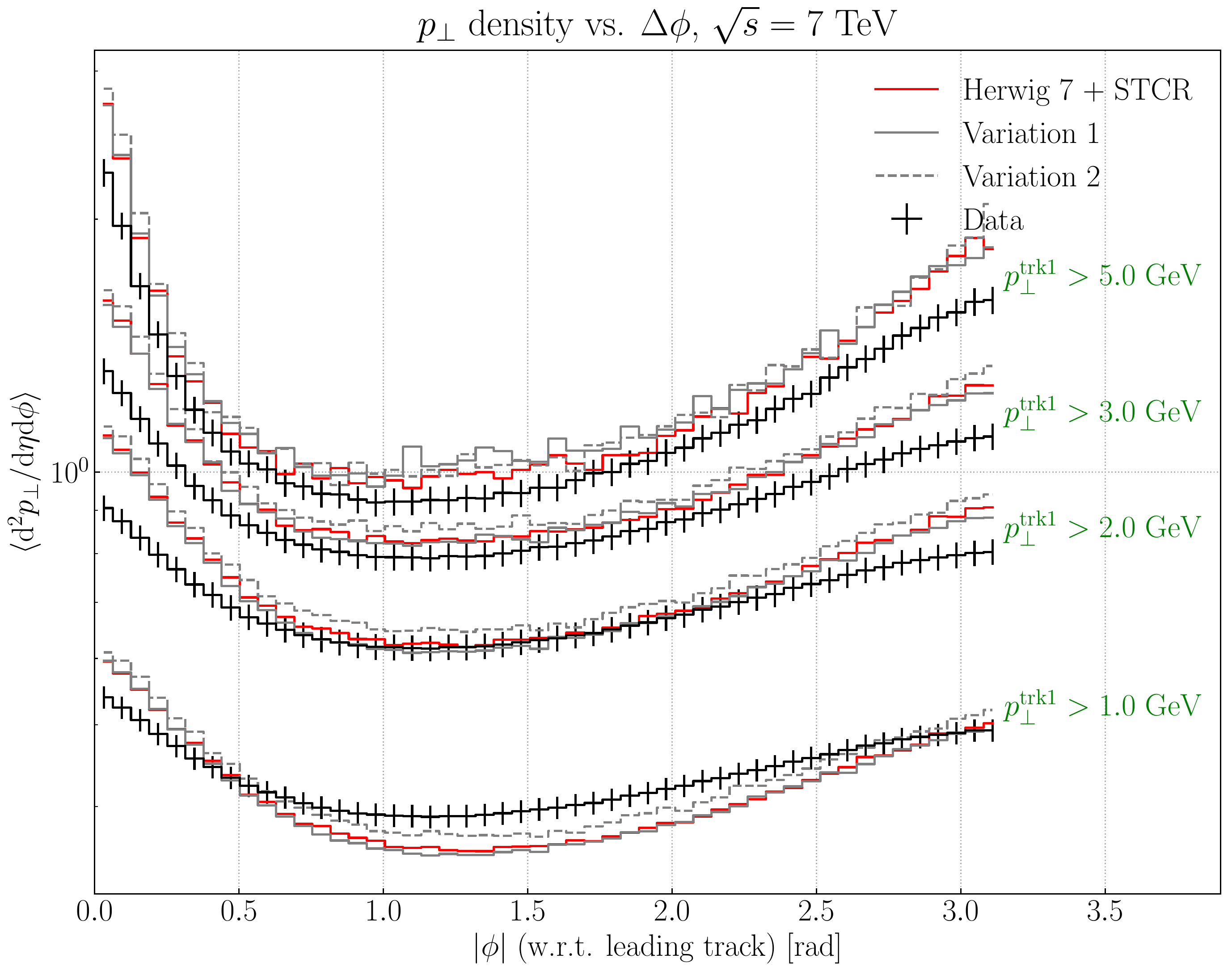}
\caption{Predictions for the average sum of particle transverse momenta as a function of 
the leading track's transverse momentum, and the average
transverse momentum as a function of the
azimuthal angle of the leading track \cite{Aad:2010fh}.}
\label{fig:Fig7}
\end{figure*}

\section{Conclusion and Outlook}
\label{sec:conclusion}
We have implemented spacetime coordinate generation for two stages
of event simulation: the positions of MPI scattering centres,
and the propagation distance in the transverse plane of partons at
the end of the parton shower. We then used these transverse coordinates
and the rapidity of the cluster constituents to define a measure that 
we minimize when performing baryonic colour
reconnection, creating a model we call baryonic spacetime colour reconnection.

Overall we find that the proposed algorithm for
baryonic spacetime colour reconnection gives meaningful 
results for many observables in Minimum Bias interactions at the LHC. 
This is an important step 
as with this prescription at hand we may explore
larger systems, where 
spacetime structure will play an important role,
as is the case in heavy ion collisions.  
However, we deliberately leave these new areas of 
study to future work after establishing the 
algorithm in $pp$ collisions in the first place.

There is plenty of room for future work based on the prescription
we present here. One avenue might be to look at only allowing
certain MPI subsystems to reconnect with each other based on
closeness in spacetime \cite{Blok:2017pui}. 
Alternatively, one may try to use the ideas
of \cite{Gieseke:2018gff} but limit the computation complexity
of the problem by only performing the soft-gluon-evolution inspired
colour reconnection in a small neighbourhood of spacetime.

One may also look to study the final state of the event in more detail
using spacetime coordinates, an avenue started by
\cite{Ferreres-Sole:2018vgo}. One interesting idea
is the interplay between Bose-Einstein correlations, and hadron position 
and extent \cite{Bialas:2013oza}.
Studying these effects could help one develop a more sophisticated
and systematic model for generating spacetime coordinates.

As perturbative calculations become more precise, improving hadronization
phenomenological models remains a key part of Monte Carlo event generator
development. Overall, we have shown that it is possible to introduce 
spacetime coordinates and then
use this information to help assist colour reconnection and potentially
other soft physics phenomena.

\section*{Acknowledgements}

The authors would like to thank Peter Skands and Mike Seymour 
for helpful comments
on the manuscript, and Boris Blok for discussions about MPI.
We thank the other members of the Herwig collaboration for input to discussions.
This work has received funding from the European
Union’s Horizon 2020 research and innovation programme 
as part of the Marie 
Sklodowska-Curie Innovative Training Network MCnetITN3 
(grant agreement no. 722104).
JB acknowledges funding by the European Research 
Council (ERC) under the European Union's Horizon 2020 
research and innovation programme, grant agreement No 668679. 
This work has been supported by the BMBF under grant number 05H18VKCC1.
CBD is supported by the Australian Government Research Training 
Program Scholarship and the J. L. William Scholarship. CBD would like to
thank Lund University for their hospitality, 
where a portion of this work was undertaken.
AS acknowledges support from the National Science
Centre, Poland Grant No. 2016/23/D/ST2/02605. MM has been supported by 
the grant 18-07846Y of the
Czech Science Foundation (GACR).

\bibliography{main}

\appendix
\section{Formation time and mean lifetime}
\label{app:time}
The discussion below is adapted from \cite{Dissertori:2003pj}.
For a branching of the kind $i\to jk$ where $j$ is the produced soft, collinear gluon,
we start with
the definition of $q_i$ and expand in terms of the products of the branching:
\begin{equation}
    \begin{split}
        q_i^2 &= \left(p_j+p_k\right)^2 \\
        &= 2p_j\cdot p_k \\
        &= 2E_jE_k\left(1-\cos\theta\right) \\
        &\sim E_jE_k\theta^2 \\
        &= \frac{E_k}{E_j}k_{\perp}^2 \\
        \mathrm{where}\;\; k_{\perp} &\coloneqq E_j\theta
    \end{split}
    \label{eq:expansion}
\end{equation}
where in the second line we have assumed the products are massless, and the fourth line
is the small angle approximation.

Using Eq. \ref{eq:spacetimePropagation} for a virtual splitting parton,
and ignoring the natural width term, one
obtains:
\begin{equation}
    \tau \sim \frac{1}{\sqrt{q_i^2}} .
    \label{eq:simpletau}
\end{equation}
Since Eq. \ref{eq:simpletau} is defined in the rest frame of the decaying parton, the
boost factor is:
\begin{equation}
    \gamma = \frac{E_i}{\sqrt{q_i^2}} = \frac{E_j+E_k}{\sqrt{q_i^2}}
\end{equation}
The lifetime in the lab frame is then:
\begin{equation}
\begin{split}
    \tau' = \gamma\tau&\sim \frac{E_j+E_k}{\sqrt{q_i^2}} \frac{1}{\sqrt{q_i^2}} \\
    &= (E_j+E_k)\frac{E_j}{E_k}\frac{1}{k_{\perp}^2} \\
    &= \frac{E_j}{k_{\perp}^2}
    \label{eq:formationtime}
\end{split}
\end{equation}
where we have used the result of Eq. \ref{eq:expansion} in the second line, and in the last
line we have used the soft approximation: $E_j \ll E_k$, i.e.  a very soft gluon produced
from a splitting where the quark takes most of the energy and momentum.

The final expression in Eq. \ref{eq:formationtime} is the standard expression
for the formation time of a massless soft, collinear gluon (see
\cite{Baier:2000mf,Casalderrey-Solana2016,Blaizot:2019muz,Dominguez:2019ges} for more details).

\end{document}